\documentclass[12pt,twoside, a4paper]{article}
\def\pd{\partial}
\def\mc{\mathcal}

\usepackage[dvips]{graphicx}
\usepackage{amssymb}
\usepackage{amssymb,amsmath}
\usepackage{graphicx}
\usepackage{dsfont}
\usepackage{caption}
\usepackage{subcaption}
\input{epsf.sty}  
\begin{document}
\begin{center}
\LARGE{\textbf{Holographic RG flows and $AdS_5$ black strings from 5D half-maximal gauged supergravity}}
\end{center}
\vspace{1 cm}
\begin{center}
\large{\textbf{H. L. Dao}$^a$ and \textbf{Parinya Karndumri}$^b$}
\end{center}
\begin{center}
$^a$Department of Physics,
National University of Singapore,
3 Science Drive 2, Singapore 117551\\
E-mail: hl.dao@u.nus.edu\\
$^b$String Theory and Supergravity Group, Department
of Physics, Faculty of Science, Chulalongkorn University, 254 Phayathai Road, Pathumwan, Bangkok 10330, Thailand \\
E-mail: parinya.ka@hotmail.com \vspace{1 cm}
\end{center}
\begin{abstract}
We study five-dimensional $N=4$ gauged supergravity coupled to five vector multiplets with compact and non-compact gauge groups $U(1)\times SU(2)\times SU(2)$ and $U(1)\times SO(3,1)$. For $U(1)\times SU(2)\times SU(2)$ gauge group, we identify $N=4$ $AdS_5$ vacua with $U(1)\times SU(2)\times SU(2)$ and $U(1)\times SU(2)_{\textrm{diag}}$ symmetries and analytically construct the corresponding holographic RG flow interpolating between these critical points. The flow describes a deformation of the dual $N=2$ SCFT driven by vacuum expectation values of dimension-two operators. In addition, we study $AdS_3\times \Sigma_2$ geometries, for $\Sigma_2$ being a two-sphere $S^2$ or a two-dimensional hyperbolic space $H^2$, dual to twisted compactifications of $N=2$ SCFTs with flavor symmetry $SU(2)$. We find a number of $AdS_3\times H^2$ solutions preserving eight supercharges for different twists from $U(1)\times U(1)\times U(1)$ and $U(1)\times U(1)_{\textrm{diag}}$ gauge fields. We numerically construct various RG flow solutions interpolating between $N=4$ $AdS_5$ critical points and these $AdS_3\times H^2$ geometries in the IR. The solutions can also be interpreted as supersymmetric black strings in asymptotically $AdS_5$ space. These types of holographic solutions are also studied in non-compact $U(1)\times SO(3,1)$ gauge group. In this case, only one $N=4$ $AdS_5$ vacuum exists, and we give an RG flow solution from this $AdS_5$ to a singular geometry in the IR corresponding to an $N=2$ non-conformal field theory. An $AdS_3\times H^2$ solution together with an RG flow between this vacuum and the $N=4$ $AdS_5$ are also given.     
\end{abstract}
\newpage
%%%%%%%%%%%%%%%%%%%%%%%%%%%%%%%%%%%%%%%%%%%%%%%%%%%%%%%%%%%%%%%%%%%%%%%%%%%%%%%%%%%%%%%%%%%%%%%%%%%%%%%%%%%%%%%%%%%%%%%%%%%%%%%%%%%%%%%%%
\section{Introduction}
AdS$_5$/CFT$_4$ correspondence has attracted much attention since the first proposal of the AdS/CFT correspondence in \cite{maldacena}. Various aspects of the very well-understood duality between type IIB theory on $AdS_5\times S^5$ and $N=4$ Super Yang-Mills (SYM) theory in four dimensions are captured by $N=8$ $SO(6)$ gauged supergravity in five dimensions which is a consistent truncation of type IIB supergravity on $S^5$ \cite{Henning_IIB_truncation}. One aspect of the AdS/CFT correspondence that has been extensively studied is holographic RG flows. There are many previous works considering these solutions both in $N=8$ five-dimensional gauged supergravity and directly in type IIB string theory, see for example \cite{GPPZ,FGPP,KW_5Dflow,N2_IIB_flow,AdS_RG_flow}.       
\\
\indent Results along this direction with less supersymmetry have also appeared in \cite{N1_SYM_fixed_point,5D_N2_DW1,Davide_5D_DW,5D_N4_flow_Davide}. In this case, gauged supergravities in five dimensions with $N<8$ supersymmetry provide a very useful framework. In this paper, we consider holographic RG flows in half-maximal $N=4$ gauged supergravity coupled to vector multiplets. This gauged supergravity has global symmetry $SO(1,1)\times SO(5,n)$, $n$ being the number of vector multiplets. Gaugings of a subgroup $G_0\subset SO(1,1)\times SO(5,n)$ have been constructed in an $SO(1,1)\times SO(5,n)$ covariant manner using the embedding tensor formalism in \cite{N4_gauged_SUGRA}, see also \cite{5D_N4_Dallagata}. The resulting solutions should describe RG flows arising from perturbing $N=2$ superconformal field theories (SCFTs) by turning on some operators or their expectation values. Holographic solutions describing these $N=2$ SCFTs and their deformations are less known compared to the $N=4$ SYM. The results of this paper will give more examples of supersymmetric RG flow solutions and should hopefully shed some light on strongly coupled dynamics of $N=2$ SCFTs. 
\\
\indent We will consider $N=4$ gauged supergravity coupled to five vector multiplets. This $N=4$ gauged supergravity has a possibility of embedding in ten dimensions since the ungauged supergravity can be obtained via a $T^5$ reduction of $N=1$ supergravity in ten dimensions similar to $N=4$ supergravity in four dimensions coupled to six vector multiplets that descends from $N=1$ ten-dimensional supergravity compactified on a $T^6$. However, it should be emphasized that the gaugings considered here have no known higher dimensional origin todate. We mainly focus on domain wall solutions interpolating between $N=4$ $AdS_5$ vacua or between an $AdS_5$ vacuum and a singular domain wall corresponding to a non-conformal field theory. These types of solutions have been extensively studied in half-maximal gauged supergravities in various space-time dimensions, see \cite{Davide_5D_DW,5D_N4_flow_Davide,7Dflow,7D_flow_Tomasiello,6Dflow,6Dflow1,tri-sasakian-flow,orbifold_flow,4Dflow,3Dflow} for an incomplete list. The solutions involve only the metric and scalar fields. 
\\
\indent We will also study solutions with some vector fields non-vanishing. These solutions interpolate between the above mentioned supersymmetric $AdS_5$ vacua and $AdS_3\times \Sigma_2$ geometries in the IR in which $\Sigma_2$ is a two-sphere ($S^2$) or a two-dimensional hyperbolic space ($H^2$). Holographically, the resulting solutions describe twisted compactifications of the dual $N=2$ SCFTs to two-dimensional SCFTs as first studied in \cite{Maldacena_Nunez_nogo}. A number of these flows across dimensions have been found within $N=8$ gauged supergravity and its truncations in \cite{Cucu_AdSD-2,BB,Wraped_D3,3D_CFT_from_LS_point,AdS3_from_5D_N8_Minwoo}, see also a universal result in \cite{flow_acrossD_bobev} and solutions obtained directly from type IIB theory in \cite{BBC}. To the best of our knowledge, solutions of this type have not appeared before in the framework of $N=4$ gauged supergravity coupled to vector multiplets, see however \cite{5D_N4_Romans} for similar solutions in pure $N=4$ gauged supergravity. Our results should give a generalization of the universal RG flows across dimensions in \cite{flow_acrossD_bobev} by turning on the twists from flavor symmetries.                                          
\\
\indent In addition, $AdS_3\times \Sigma_2$ geometries can arise as near horizon limits of black strings. Therefore, flow solutions interpolating between $AdS_5$ and $AdS_3\times \Sigma_2$ should describe black strings in asymptotically $AdS_5$ space. Similar solutions in $N=2$ gauged supergravity have been considered in \cite{black_string_Klemm1,black_string_Klemm2,black_string_Klemm3,black_string_Klemm4,black_string_Klemm5}. We will give solutions of this type in $N=4$ gauged supergravity. The solutions presented here will provide further examples of supersymmetric $AdS_5$ black strings and might be useful for both holographic studies of twisted $N=2$ SCFTs on $\Sigma_2$ and certain dynamical aspects of black strings.  
\\
\indent The paper is organized as follow. In section \ref{N4_SUGRA},
we review $N=4$ gauged supergravity in five dimensions coupled to vector multiplets using the embedding tensor formalism. In section \ref{U1_SU2_SU2_gauge_group}, a compact $U(1)\times SU(2)\times SU(2)$ gauge group is considered. Supersymmetric $AdS_5$ vacua and RG flows interpolating between them are given. A number of $AdS_3\times H^2$ solutions will also be given along with numerical RG flows interpolating between the previously identified $AdS_5$ vacua and these $AdS_3\times H^2$ geometries. In section \ref{U1_SO3_1_gauge_group}, we repeat the analysis for a non-compact $U(1)\times SO(3,1)$ gauge group. An RG flow from $N=2$ SCFT dual to a supersymmetric $AdS_5$ vacuum to a singular geometry dual to a non-conformal field theory is considered. A supersymmetric $AdS_3\times H^2$ geometry and an RG flow between this vacuum and the $AdS_5$ critical point will also be given. We end the paper with some conclusions and comments in section \ref{conclusion}.

%%%%%%%%%%%%%%%%%%%%%%%%%%%%%%%%%%%%%%%%%%%%%%%%%%%%%%%%%%%%%%%%%%%%%%%%%%%%%%%%%%%%%%%%%%%%%%%%%%%%%%%%%%%%%%%%%%%%%%%%%%%%%%%%%%%%%%%%%
\section{Five dimensional $N=4$ gauged supergravity coupled to vector multiplets}\label{N4_SUGRA} 
In this section, we review the structure of five dimensional $N=4$ gauged supergravity coupled to vector multiplets. We mainly focus on relevant formulae to find supersymmetric solutions. More details on the construction of $N=4$ gauged supergravity can be found in \cite{N4_gauged_SUGRA} and \cite{5D_N4_Dallagata}. 
\\
\indent In five dimensions, $N=4$ gravity multiplet consists of the graviton
$e^{\hat{\mu}}_\mu$, four gravitini $\psi_{\mu i}$, six vectors $A^0$ and
$A_\mu^m$, four spin-$\frac{1}{2}$ fields $\chi_i$ and one real
scalar $\Sigma$, the dilaton. Space-time and tangent space indices are denoted respectively by $\mu,\nu,\ldots =0,1,2,3,4$ and
$\hat{\mu},\hat{\nu},\ldots=0,1,2,3,4$. The $SO(5)\sim USp(4)$
R-symmetry indices are described by $m,n=1,\ldots, 5$ for the
$SO(5)$ vector representation and $i,j=1,2,3,4$ for the $SO(5)$
spinor or $USp(4)$ fundamental representation.
\\
\indent $N=4$ supersymmetry allows the gravity multiplet to couple to an arbitrary number $n$ of vector multiplets. Each vector multiplet contains a vector field $A_\mu$, four gaugini $\lambda_i$ and five scalars $\phi^m$. The $n$ vector
multiplets will be labeled by indices $a,b=1,\ldots, n$. Components fields in the $n$ vector multiplets are accordingly denoted by $(A^a_\mu,\lambda^{a}_i,\phi^{ma})$. The $5n$ scalar fields parametrized the $SO(5,n)/SO(5)\times SO(n)$ coset. Combining the gravity and vector multiplets, we have $6+n$ vector fields denoted by $A^{\mc{M}}_\mu=(A^0_\mu,A^m_\mu,A^a_\mu)$ and $5n+1$ scalars. All fermionic fields are symplectic Majorana spinors subject to the condition
\begin{equation}
\xi_i=\Omega_{ij}C(\bar{\xi}^j)^T 
\end{equation}
with $C$ and $\Omega_{ij}$ being the charge conjugation matrix and $USp(4)$ symplectic form, respectively.
\\
\indent To describe the $SO(5,n)/SO(5)\times SO(n)$ coset, we introduce a coset representative $\mc{V}_M^{\phantom{M}A}$ transforming under the global $G=SO(5,n)$ and the local $H=SO(5)\times SO(n)$ by left and right multiplications, respectively. We use indices $M,N,\ldots,=1,2,\ldots , 5+n$. The local $H$ indices $A,B,\ldots$ can be split into $A=(m,a)$. We can then write the coset representative as
\begin{equation}
\mc{V}_M^{\phantom{M}A}=(\mc{V}_M^{\phantom{M}m},\mc{V}_M^{\phantom{M}a}).
\end{equation}
The matrix $\mc{V}_M^{\phantom{M}A}$, an element of $SO(5,n)$, satisfies the relation
\begin{equation}
\eta_{MN}={\mc{V}_M}^A{\mc{V}_N}^B\eta_{AB}=-\mc{V}_M^{\phantom{M}m}\mc{V}_N^{\phantom{M}m}+\mc{V}_M^{\phantom{M}a}\mc{V}_N^{\phantom{M}a}\,
.
\end{equation}
$\eta_{MN}=\textrm{diag}(-1,-1,-1,-1,-1,1,\ldots,1)$ is the $SO(5,n)$ invariant tensor. In addition, the $SO(5,n)/SO(5)\times SO(n)$ coset can also be described in term of a symmetric matrix
\begin{equation}
M_{MN}=\mc{V}_M^{\phantom{M}m}\mc{V}_N^{\phantom{M}m}+\mc{V}_M^{\phantom{M}a}\mc{V}_N^{\phantom{M}a}
\end{equation}
which is manifestly invariant under the $SO(5)\times SO(n)$  local symmetry.
\\
\indent The full global symmetry of $N=4$ supergravity coupled to $n$ vector multiplets is $SO(1,1)\times SO(5,n)$. The $SO(1,1)\sim \mathbb{R}^+$ factor is identified with the coset space described by the dilaton $\Sigma$. Gaugings can be efficiently described, in an $SO(1,1)\times SO(5,n)$ covariant manner, by using the embedding tensor formalism. $N=4$ supersymmetry allows three components of the embedding tensor $\xi^{M}$, $\xi^{MN}=\xi^{[MN]}$ and $f_{MNP}=f_{[MNP]}$. The existence of supersymmetric $AdS_5$ vacua requires $\xi^M=0$, see \cite{AdS5_N4_Jan} for more detail. Since, in this paper, we are only interested in supersymmetric $AdS_5$ vacua and solutions interpolating between these vacua or solutions asymptotically approaching $AdS_5$, we will restrict ourselves to the gaugings with $\xi^{M}=0$.
\\
\indent With $\xi^{M}=0$, the gauge group is entirely embedded in $SO(5,n)$. The gauge generators in the fundamental representation of $SO(5,n)$ can be written in terms of the $SO(5,n)$ generators ${(t_{MN})_P}^Q=\delta^Q_{[M}\eta_{N]P}$ as
\begin{equation}
{(X_M)_N}^P=-{f_M}^{QR}{(t_{QR})_N}^P={f_{MN}}^P\quad \textrm{and}\quad {(X_0)_N}^P=-\xi^{QR}{(t_{QR})_N}^P={\xi_N}^P\, .
\end{equation}
As a result, the covariant derivative reads
\begin{equation}
D_\mu=\nabla_\mu+A_\mu^{M}X_M+A^0_\mu X_0
\end{equation}
where $\nabla_\mu$ is the usual space-time covariant derivative including the spin connection. It should be noted that the definition of $\xi^{MN}$ and $f_{MNP}$ includes the gauge coupling constants. Furthermore, $SO(5,n)$ indices $M,N,\ldots$ are lowered and raised by $\eta_{MN}$ and its inverse $\eta^{MN}$. 
\\
\indent In order to define a consistent gauge group, generators $X_{\mc{M}}=(X_0,X_M)$ must form a closed subalgebra of $SO(5,n)$. This requires $\xi^{MN}$ and $f_{MNP}$ to satisfy the quadratic constraints
\begin{equation}
f_{R[MN}{f_{PQ]}}^R=0\qquad \textrm{and}\qquad {\xi_M}^Qf_{QNP}=0\, .
\end{equation}      
The first condition is the usual Jacobi identity. From the result of \cite{AdS5_N4_Jan}, gauge groups that admit $N=4$ supersymmetric $AdS_5$ vacua are generally of the form $U(1)\times H_0\times H$. The $U(1)$ is gauged by $A^0_\mu$ while $H\subset SO(n+3-\textrm{dim}\, H_0)$ is a compact group gauged by vector fields in the vector multiplets. $H_0$ is a non-compact group gauged by three of the graviphotons and $\textrm{dim}\, H_0-3$ vectors from the vector multiplets. In addition, $H_0$ must contain an $SU(2)$ subgroup. For simple groups, $H_0$ can be $SU(2)\sim SO(3)$, $SO(3,1)$ and $SL(3,\mathbb{R})$.  
\\
\indent The bosonic Lagrangian of a general gauged $N=4$ supergravity coupled to $n$ vector multiplets can be written as
\begin{eqnarray}
e^{-1}\mc{L}&=&\frac{1}{2}R-\frac{1}{4}\Sigma^2M_{MN}\mc{H}^M_{\mu\nu}\mc{H}^{N\mu\nu}-\frac{1}{4}\Sigma^{-4}\mc{H}^0_{\mu\nu}\mc{H}^{0\mu\nu}\nonumber \\
& &-\frac{3}{2}\Sigma^{-2}D_\mu \Sigma D^\mu \Sigma +\frac{1}{16} D_\mu M_{MN}D^\mu
M^{MN}-V+e^{-1}\mc{L}_{\textrm{top}}
\end{eqnarray}
where $e$ is the vielbein determinant. $\mc{L}_{\textrm{top}}$ is the topological term which we will not give the explicit form here due to its complexity. The covariant gauge field strength tensors read
\begin{equation}
\mc{H}^{\mc{M}}_{\mu\nu}=2\pd_{[\mu}A^{\mc{M}}_{\nu]}+{X_{\mc{N}\mc{P}}}^{\mc{M}}A^{\mc{N}}_\mu A^{\mc{P}}_\nu+Z^{\mc{M}\mc{N}}B_{\mc{N}\mu\nu}
\end{equation}
where 
\begin{eqnarray}
{X_{MN}}^P&=&-{f_{MN}}^P,\qquad {X_{M0}}^0=0,\qquad {X_{0M}}^N=-{\xi_M}^N,\nonumber \\
 Z^{MN}&=&\frac{1}{2}\xi^{MN},\qquad Z^{0M}=-Z^{M0}=0\, . 
\end{eqnarray}
The two-form fields do not have kinetic terms and satisfy the first-order field equation
\begin{equation}
Z^{\mc{M}\mc{N}}\left[\frac{1}{6\sqrt{2}}\epsilon_{\mu\nu\rho\lambda\sigma}\mc{H}^{(3)\rho\lambda\sigma}_{\mc{N}}-\mc{M}_{\mc{N}\mc{P}}
\mc{H}^{\mc{P}}_{\mu\nu}\right]=0
\end{equation}
with $\mc{H}^{(3)}$ defined by
\begin{equation}
Z^{\mc{M}\mc{N}}\mc{H}^{(3)}_{\mc{N}\mu\nu\rho}=Z^{\mc{M}\mc{N}}\left[3D_{[\mu}B_{\nu\rho]\mc{N}}
+6d_{\mc{NPQ}}A^{\mc{P}}_{[\mu}\left(\pd_\nu A^{\mc{Q}}_{\rho]}+\frac{1}{3}{X_{\mc{RS}}}^{\mc{Q}}A^{\mc{R}}_\nu A^{\mc{S}}_{\rho]}\right)\right]
\end{equation}
and $d_{0MN}=d_{MN0}=d_{M0N}=\eta_{MN}$. These two form fields arise from vector fields that transform non-trivially under the $U(1)$ part of the gauge group.
\\
\indent The scalar potential is given by
\begin{eqnarray}
V&=&-\frac{1}{4}\left[f_{MNP}f_{QRS}\Sigma^{-2}\left(\frac{1}{12}M^{MQ}M^{NR}M^{PS}-\frac{1}{4}M^{MQ}\eta^{NR}\eta^{PS}\right.\right.\nonumber \\
& &\left.+\frac{1}{6}\eta^{MQ}\eta^{NR}\eta^{PS}\right) +\frac{1}{4}\xi_{MN}\xi_{PQ}\Sigma^4(M^{MP}M^{NQ}-\eta^{MP}\eta^{NQ})\nonumber \\
& &\left.
+\frac{\sqrt{2}}{3}f_{MNP}\xi_{QR}\Sigma M^{MNPQRS}\right]
\end{eqnarray}
where $M^{MN}$ is the inverse of $M_{MN}$, and $M^{MNPQRS}$ is obtained from
\begin{equation}
M_{MNPQR}=\epsilon_{mnpqr}\mc{V}_{M}^{\phantom{M}m}\mc{V}_{N}^{\phantom{M}n}
\mc{V}_{P}^{\phantom{M}p}\mc{V}_{Q}^{\phantom{M}q}\mc{V}_{R}^{\phantom{M}r}
\end{equation}
by raising the indices with $\eta^{MN}$. 
\\
\indent Fermionic supersymmetry transformations are given by
\begin{eqnarray}
\delta\psi_{\mu i} &=&D_\mu \epsilon_i+\frac{i}{\sqrt{6}}\Omega_{ij}A^{jk}_1\gamma_\mu\epsilon_k\nonumber \\
& &-\frac{i}{6}\left(\Omega_{ij}\Sigma{\mc{V}_M}^{jk}\mc{H}^M_{\nu\rho}-\frac{\sqrt{2}}{4}\delta^k_i\Sigma^{-2}\mc{H}^0_{\nu\rho}\right)({\gamma_\mu}^{\nu\rho}-4\delta^\nu_\mu\gamma^\rho)\epsilon_k,\\
\delta \chi_i &=&-\frac{\sqrt{3}}{2}i\Sigma^{-1} D_\mu
\Sigma\gamma^\mu \epsilon_i+\sqrt{2}A_2^{kj}\epsilon_k\nonumber \\
& &-\frac{1}{2\sqrt{3}}\left(\Sigma \Omega_{ij}{\mc{V}_M}^{jk}\mc{H}^M_{\mu\nu}+\frac{1}{\sqrt{2}}\Sigma^{-2}\delta^k_i\mc{H}^0_{\mu\nu}\right)\gamma^{\mu\nu}\epsilon_k,\\
\delta \lambda^a_i&=&i\Omega^{jk}({\mc{V}_M}^aD_\mu
{\mc{V}_{ij}}^M)\gamma^\mu\epsilon_k+\sqrt{2}\Omega_{ij}A_{2}^{akj}\epsilon_k-\frac{1}{4}\Sigma{\mc{V}_M}^a\mc{H}^M_{\mu\nu}\gamma^{\mu\nu}\epsilon_i\,
.
\end{eqnarray}
In the above equations, the fermion shift matrices are defined by
\begin{eqnarray}
A_1^{ij}&=&-\frac{1}{\sqrt{6}}\left(\sqrt{2}\Sigma^2\Omega_{kl}{\mc{V}_M}^{ik}{\mc{V}_N}^{jl}\xi^{MN}+\frac{4}{3}\Sigma^{-1}{\mc{V}^{ik}}_M{\mc{V}^{jl}}_N{\mc{V}^P}_{kl}{f^{MN}}_P\right),\nonumber
\\
A_2^{ij}&=&\frac{1}{\sqrt{6}}\left(\sqrt{2}\Sigma^2\Omega_{kl}{\mc{V}_M}^{ik}{\mc{V}_N}^{jl}\xi^{MN}-\frac{2}{3}\Sigma^{-1}{\mc{V}^{ik}}_M{\mc{V}^{jl}}_N{\mc{V}^P}_{kl}{f^{MN}}_P\right),\nonumber
\\
A_2^{aij}&=&-\frac{1}{2}\left(\Sigma^2{\mc{V}_M}^a{\mc{V}_N}^{ij}\xi^{MN}-\sqrt{2}\Sigma^{-1}\Omega_{kl}{\mc{V}_M}^a{\mc{V}_N}^{ik}{\mc{V}_P}^{jl}f^{MNP}\right).
\end{eqnarray}
\indent $\mc{V}_M^{\phantom{M}ij}$ is defined in term of ${\mc{V}_M}^m$ as
\begin{equation}
{\mc{V}_M}^{ij}=\frac{1}{2}{\mc{V}_M}^{m}\Gamma^{ij}_m
\end{equation}
where $\Gamma^{ij}_m=\Omega^{ik}{\Gamma_{mk}}^j$ and ${\Gamma_{mi}}^j$ are $SO(5)$ gamma matrices. Similarly, the inverse ${\mc{V}_{ij}}^M$ can be written as
\begin{equation}
{\mc{V}_{ij}}^M=\frac{1}{2}{\mc{V}_m}^M(\Gamma^{ij}_m)^*=\frac{1}{2}{\mc{V}_m}^M\Gamma_{m}^{kl}\Omega_{ki}\Omega_{lj}\,
.
\end{equation}
The covariant derivative on $\epsilon_i$ is given by
\begin{equation}
D_\mu \epsilon_i=\pd_\mu \epsilon_i+\frac{1}{4}\omega_\mu^{ab}\gamma_{ab}\epsilon_i+{Q_{\mu i}}^j\epsilon_j
\end{equation}
where the composite connection is defined by
\begin{equation}
{Q_{\mu i}}^j={\mc{V}_{ik}}^M\pd_\mu {\mc{V}_M}^{kj}-A^0_\mu\xi^{MN}\mc{V}_{Mik}{\mc{V}_N}^{kj}-A^M_\mu{\mc{V}_{ik}}^N\mc{V}^{kjP}f_{MNP}\, .
\end{equation}
\indent Before considering various supersymmetric solutions, we note here the relation between the scalar potential and the fermion shift matrices $A_1$ and $A_2$
\begin{equation}
V=-A^{ij}_1A_{1ij}+A^{ij}_2A_{2ij}+A_2^{aij}
A^a_{2ij}\, .
\end{equation}
Rasing and lowering of indices $i,j,\ldots$ by $\Omega^{ij}$ and $\Omega_{ij}$ are also related to complex conjugation for example $A_{1ij}=\Omega_{ik}\Omega_{jl}A_1^{kl}=(A_1^{ij})^*$.

%%%%%%%%%%%%%%%%%%%%%%%%%%%%%%%%%%%%%%%%%%%%%%%%%%%%%%%%%%%%%%%%%%%%%%%%%%%%%%%%%%%%%%%%%%%%%%%%%%%%%%%%%%%%%%%%%%%%%%%%%%%%%%%%%%%%%%%%%
\section{Supersymmetric RG flows in $U(1)\times SU(2)\times SU(2)$ gauge group}\label{U1_SU2_SU2_gauge_group}
We begin with a compact gauge group $U(1)\times SU(2)\times SU(2)$. In order to gauge this group, we need to couple the gravity multiplet to at least three vector multiplets. Components of the embedding tensor for this gauge group are given by
\begin{eqnarray}
\xi^{MN}&=&g_1(\delta^M_2\delta^N_1-\delta^M_1\delta^N_2),\\ 
f_{\tilde{m}+2,\tilde{n}+2,\tilde{p}+2}&=&-g_2\epsilon_{\tilde{m}\tilde{n}\tilde{p}},\qquad \tilde{m},\tilde{n},\tilde{p}=1,2,3,\\
f_{abc}&=&g_3\epsilon_{abc},\qquad a,b,c=1,2,3
\end{eqnarray} 
where $g_1$, $g_2$ and $g_3$ are the corresponding coupling constants for each factor in $U(1)\times SU(2)\times SU(2)$.
\\
\indent To parametrize the scalar coset $SO(5,n)/SO(5)\times SO(n)$, we introduce a basis for $GL(5+n,\mathbb{R})$ matrices
\begin{equation} 
(e_{MN})_{PQ}=\delta_{MP}\delta_{NQ}
\end{equation}
in terms of which $SO(5,n)$ non-compact generators are given by
\begin{equation}
Y_{ma}=e_{m,a+5}+e_{a+5,m},\qquad m=1,2,\ldots, 5,\qquad a=1,2,\ldots, n\, .
\end{equation}
We will mainly consider the case of $n=5$ vector multiplets, but the results can be straightforwardly extended to the case of $n>5$. 

\subsection{RG flows between $N=4$ supersymmetric $AdS_5$ critical points}
We will consider scalar fields that are singlets of $U(1)\times SU(2)_{\textrm{diag}}\subset U(1)\times SU(2)\times SU(2)$. Under $SO(5)\times SO(5)\subset SO(5,5)$, the scalars transform as $(\mathbf{5},\mathbf{5})$. With the above embedding of the gauge group in $SO(5,5)$, the scalars transform under $U(1)\times SU(2)\times SU(2)$ gauge group as
\begin{equation}
2\times (\mathbf{1},\mathbf{1})_{+2}+2\times (\mathbf{1},\mathbf{1})_{-2}+(\mathbf{1},\mathbf{3})_{+2}+(\mathbf{1},\mathbf{3})_{-2}+2\times (\mathbf{3},\mathbf{1})_{0}+(\mathbf{3},\mathbf{3})_{0}
\end{equation}  
and transform under $U(1)\times SU(2)_{\textrm{diag}}$ as
\begin{equation}  
\mathbf{1}_0+2\times \mathbf{1}_{+2}+2\times \mathbf{1}_{-2}+3\times \mathbf{3}_0+\mathbf{3}_{+2}+\mathbf{3}_{-2}+\mathbf{5}_0
\end{equation}
where the subscript denotes the $U(1)$ charges. According to this decomposition, there is one singlet corresponding to the following $SO(5,5)$ non-compact generator
\begin{equation}
Y_s=Y_{31}+Y_{42}+Y_{53}\, .
\end{equation}
Using the coset representative parametrized by
\begin{equation}
\mc{V}=e^{\phi Y_s},\label{U1_SU2d_coset}
\end{equation}
we find the scalar potential for $\phi$ and $\Sigma$ as follow
\begin{eqnarray}
V&=&\frac{1}{32\Sigma^2}\left[32\sqrt{2}g_1g_2\Sigma^3\cosh^3\phi-9(g_2^2+g_3^2)\cosh(2\phi) \right.\nonumber \\
& &-8(g_2^2-g_3^2-4\sqrt{2}g_1g_3\Sigma^3\sinh^3\phi-g_2g_3\sinh^3\phi)\nonumber \\ 
& &\left.+(g_2^2+g_3^2)\cosh(6\phi)\right].\label{potential_SU2d}
\end{eqnarray}
This potential admits two $N=4$ supersymmetric $AdS_5$ critical points. The first one is given by
\begin{equation}
\phi=0\qquad \textrm{and}\qquad \Sigma=\left(-\frac{g_2}{\sqrt{2}g_1}\right)^{\frac{1}{3}}\, .\label{AdS5_1_compact}
\end{equation}
This critical point is invariant under the full gauge symmetry $U(1)\times SU(2)\times SU(2)$ since $\Sigma$ is a singlet of the whole $SO(5,5)$ global symmetry. Furthermore, we can rescale $\Sigma$, or equivalently set $g_2=-\sqrt{2}g_1$ to bring this critical point located at $\Sigma=1$. The cosmological constant, the value of the scalar potential at the critical point, is 
\begin{equation}
V_0=-3g_1^2\, .
\end{equation}    
\indent Another supersymmetric critical point is given by 
\begin{equation}
\phi=\frac{1}{2}\ln\left[\frac{g_3-g_2}{g_3+g_2}\right]\qquad \textrm{and}\qquad \Sigma=\left(\frac{g_2g_3}{g_1\sqrt{2(g_3^2-g_2^2)}}\right)^{\frac{1}{3}}\, .\label{AdS5_2_compact}
\end{equation}
This critical point also preserves the full $N=4$ supersymmetry but has only $U(1)\times SU(2)_{\textrm{diag}}$ symmetry due to the non-vanising scalar $\phi$. The cosmological constant for this critical point is 
\begin{equation}
V_0=-3\left(\frac{g_1g_2^2g_3^2}{2(g_3^2-g_2^2)}\right)^{\frac{2}{3}}\, .
\end{equation}
This second $N=4$ $AdS_5$ critical point has been shown to exist in \cite{5D_N4_flow_Davide} when an additional $SU(2)$ dual to a flavor symmetry of the dual $N=2$ SCFT is present.
\\
\indent We now analyze the BPS equations arising from setting supersymmetry transformations of fermions to zero. We first define ${\mc{V}_M}^{ij}$ with the following explicit choice of $SO(5)$ gamma matrices ${\Gamma_{mi}}^j$
\begin{eqnarray}
\Gamma_1&=&-\sigma_2\otimes \sigma_2,\qquad \Gamma_2=i\mathbb{I}_2\otimes \sigma_1,\qquad \Gamma_3=\mathbb{I}_2\otimes \sigma_3,\nonumber\\
\Gamma_4&=&\sigma_1\otimes \sigma_2,\qquad \Gamma_5=\sigma_3\otimes \sigma_2
\end{eqnarray}
where $\sigma_i$, $i=1,2,3$ are the usual Pauli matrices.
\\
\indent Since we are interested only in solutions with only the metric and scalars non-vanishing, we will set all the vector and two-form fields to zero. The metric ansatz is given by the standard domain wall
\begin{equation}
ds^2=e^{2A(r)}dx^2_{1,3}+dr^2
\end{equation} 
with $dx^2_{1,3}$ being the metric on four dimensional Minkowski space. In addition, the scalars $\Sigma$ and $\phi$ as well as the Killing spinors $\epsilon_i$ are functions of only the radial coordinate $r$.
\\
\indent We begin with the $\delta \psi_{\mu i}=0$ conditions for $\mu=0,1,2,3$ which lead to
\begin{equation}
A'\gamma_r\epsilon_i+i\sqrt{\frac{2}{3}}\Omega_{ij}A^{jk}_1\epsilon_k=0\label{psi_eq1}
\end{equation}
where $'$ denotes the $r$-derivative. Multiply this equation by $A'\gamma_r$ and iterate, we find
\begin{equation}
A'^2\epsilon_i+{M_i}^k{M_k}^j\epsilon_j=0\label{psi_eq2}
\end{equation}
for ${M_i}^j=\sqrt{\frac{2}{3}}\Omega_{ik}A_1^{kj}$. The above equation has non-vanishing solutions for $\epsilon_i$ if ${M_i}^k{M_k}^j\propto \delta^j_i$. We will write
\begin{equation}
{M_i}^k{M_k}^j=-|W|^2\delta^j_i
\end{equation}
where $W$ will be identified with the superpotential. When substitute this result in equation \eqref{psi_eq2}, we find
\begin{equation}
A'=\pm |W|\, .
\end{equation}
On the other hand, equation \eqref{psi_eq1} leads to the projection condition on $\epsilon_i$
\begin{equation}
\gamma_r\epsilon_i=\pm i {I_i}^j\epsilon_j\label{gamma_r_projector}
\end{equation} 
where ${I_i}^j$ is defined via
\begin{equation}
{M_i}^j=|W|{I_i}^j\, .
\end{equation}
The condition $\delta \psi_{\hat{r}i}=0$ gives the usual $r$-dependent Killing spinors of the form $\epsilon_i=e^{\frac{A}{2}}\epsilon_{0i}$ for constant spinors $\epsilon_{0i}$ satisfying \eqref{gamma_r_projector}. Using the projector \eqref{gamma_r_projector} in conditions $\delta\chi_i=0$ and $\delta \lambda^a_i=0$, we can derive the first order flow equations for $\Sigma$ and $\phi$. 
\\
\indent 
For the coset representative in \eqref{U1_SU2d_coset}, we find the superpotential
\begin{equation}
W=\frac{1}{6}\Sigma^{-1}\left(\sqrt{2}g_1\Sigma^3-2g_2\cosh^3\phi-2g_3\sinh^3\phi\right).
\end{equation}
The matrix ${I_i}^j$ in the $\gamma_r$ projection is given by
\begin{equation} 
i{I_i}^j={(\sigma_2\otimes \sigma_3)_i}^j\, .\label{I_gamma_r}
\end{equation}
The scalar kinetic term reads
\begin{equation}
\mc{L}_{\textrm{kin}}=-\frac{3}{2}\Sigma^{-2}\Sigma'^2-\frac{3}{2}\phi'^2 \, .
\end{equation}
The scalar potential \eqref{potential_SU2d} can be written in term of the superpotential as
\begin{equation}
V=\frac{3}{2}\left(\frac{\pd W}{\pd \phi}\right)^2+\frac{3}{2}\Sigma^2\left(\frac{\pd W}{\pd \Sigma}\right)^2-6W^2\, .
\end{equation}
\indent By choosing the sign choice such that the $U(1)\times SU(2)\times SU(2)$ is identified with $r\rightarrow \infty$, the BPS conditions from $\delta\chi_i$ and $\delta\lambda^a_i$ reduce to the following equations 
\begin{eqnarray}
\phi'&=&-\frac{\pd W}{\pd \phi}=\Sigma^{-1}\cosh\phi\sinh\phi(g_2\cosh\phi+g_3\sinh\phi),\label{phi_eq1}\\
\Sigma'&=&-\Sigma^2\frac{\pd W}{\pd \Sigma}=-\frac{1}{3}(\sqrt{2}g_1\Sigma^3+g_2\cosh^3\phi+g_3\sinh^3\phi).\label{Sigma_eq1}
\end{eqnarray}
\indent It can be readily verified that the critical points given in \eqref{AdS5_1_compact} and \eqref{AdS5_2_compact} are also critical point of $W$ and solve equations \eqref{phi_eq1} and \eqref{Sigma_eq1}. These critical points are then $N=4$ supersymmetric. Together with the $A'$ equation
\begin{equation}
A'=-\frac{1}{6}\Sigma^{-1}(\sqrt{2}g_1\Sigma^3-2g_2\cosh^3\phi-2g_3\sinh^3\phi),
\end{equation}
we have the full set of BPS equations to be solved for RG flows interpolating between the two supersymmetric $AdS_5$ critical points. It can be verified that these BPS equations imply the second-order field equations.
\\
\indent By treating $\phi$ as an independent variable, we can solve for $\Sigma(\phi)$ and $A(\phi)$ as follow
\begin{eqnarray}
\Sigma&=&-\frac{e^{\frac{\phi}{3}}(g_3-g_3e^{2\phi}-g_2-g_2e^{2\phi})^{\frac{1}{3}}}{\left[(e^{4\phi}-1)C-2\sqrt{2}g_1\right]^{\frac{1}{3}}},\\
A&=&-\frac{1}{3}\phi+\frac{1}{2}\ln(1-e^{4\phi})-\frac{1}{2}\ln(g_3-g_3e^{2\phi}-g_2-g_2e^{2\phi})\nonumber \\
& &-\frac{1}{6}\ln\left[g_2(1-e^{2\phi})-g_3(1+e^{2\phi})\right].
\end{eqnarray}
We have neglected an irrelevant additive integration constant in $A$. The constant $C$ will be chosen in such a way that $\Sigma$ approaches the second $AdS_5$ vacuum. This requires $C=-\frac{g_1(g_3+g_2)^2}{\sqrt{2}g_2g_3}$ leading to the final form of the solution for $\Sigma$  
\begin{equation}
\Sigma=\left[\frac{\sqrt{2}g_2g_3e^{\phi}}{g_1(g_2-g_2e^{2\phi}-g_3-g_3e^{2\phi})}\right]^{\frac{1}{3}}\, .
\end{equation}
Finally, the solution for $\phi(r)$ is given by
\begin{equation}
g_2g_3\rho=g_3\ln\left[\frac{1-e^\phi}{1+e^\phi}\right]-2g_2\tan^{-1}e^\phi+2\sqrt{g_3^2-g_2^2}\tanh^{-1}\left[e^\phi\sqrt{\frac{g_3+g_2}{g_3-g_2}}\right]
\end{equation}
where the new radial coordinate $\rho$ is defined by $\frac{d\rho}{dr}=\Sigma^{-1}$. This solution is the same as that obtained in \cite{5D_N4_flow_Davide} and has a very similar structure to solutions obtained from half-maximal gauged supergravities in seven and six dimensions \cite{7Dflow,6Dflow}.
\\
\indent Near the UV $N=4$ critical point, we find
\begin{equation}
\phi\sim \Sigma\sim e^{-\sqrt{2}g_1r}\sim e^{-\frac{2r}{L_{\textrm{UV}}}}
\end{equation}
where the $AdS_5$ radius is given by $L_{\textrm{UV}}=\sqrt{-\frac{6}{V_0}}=\frac{\sqrt{2}}{g_1}$. This behavior implies that the RG flow dual to this solution is driven by vacuum expectation values of operators with dimension $\Delta=2$. Near the IR point, we find
\begin{equation}
\phi\sim e^{\frac{2r}{L_{\textrm{IR}}}},\qquad \Sigma\sim e^{-\frac{2r}{L_{\textrm{IR}}}}
\end{equation} 
where 
\begin{equation}
L_{\textrm{IR}}=\left[\frac{2^{\frac{5}{2}}(g_3^2-g_2^2)}{g_1(g_2g_3)^2}\right]^{\frac{1}{3}}\, .
\end{equation}
The operator dual to $\phi$ becomes irrelevant in the IR with dimension $\Delta=6$ while the operator dual to $\Sigma$ has dimension $\Delta=2$ as in the UV. For completeness, we give masses for all scalars at both critical points in table \ref{table1} and \ref{table2}. In these tables, the singlets $(\mathbf{1},\mathbf{1})_0$ and one of the $\mathbf{1}_0$ with $m^2L^2=-4$ in table \ref{table2} corresponds to $\Sigma$. The massless scalars $\mathbf{3}_0$ in table \ref{table2} are Goldstone bosons corresponding to the symmetry breaking $SU(2)\times SU(2)\rightarrow SU(2)_{\textrm{diag}}$. The massless scalars $\mathbf{5}_0$ are dual to marginal operators in the dual $N=2$ SCFT. It should also be noted that most of the results in this section have already been found in \cite{5D_N4_flow_Davide}. However, the full scalar mass spectra are new results that have not been studied in \cite{5D_N4_flow_Davide}. 
\\
\begin{table}[h]
\centering
\begin{tabular}{|c|c|c|}
  \hline
  Scalar field representations & $m^2L^2\phantom{\frac{1}{2}}$ & $\Delta$  \\ \hline
  $(\mathbf{1},\mathbf{1})_0$ & $-4$ &  $2$  \\
  $(\mathbf{1},\mathbf{1})_{\pm 2}$ & $-3_{\times 4}$ &  $3$  \\
   $(\mathbf{1},\mathbf{3})_{\pm 2}$ & $-3_{\times 6}$ &  $3$  \\
    $(\mathbf{3},\mathbf{1})_0$ & $-4_{\times 6}$ &  $2$  \\
      $(\mathbf{3},\mathbf{3})_0$ & $-4_{\times 9}$ &  $2$  \\
  \hline
\end{tabular}
\caption{Scalar masses at the $N=4$ supersymmetric $AdS_5$ critical
point with $U(1)\times SU(2)\times SU(2)$ symmetry and the
corresponding dimensions of the dual operators.}\label{table1}
\end{table}

\begin{table}[h]
\centering
\begin{tabular}{|c|c|c|}
  \hline
  Scalar field representations & $m^2L^2\phantom{\frac{1}{2}}$ & $\Delta$  \\ \hline
  $\mathbf{1}_0$ & $-4$ &  $2$  \\
  $\mathbf{1}_0$ & $12$ &  $6$  \\
  $\mathbf{1}_{\pm 2}$ & $-3_{\times 4}$ &  $3$  \\
   $\mathbf{3}_{\pm 2}$ & $5_{\times 6}$ &  $5$  \\
    $\mathbf{3}_0$ & $-4_{\times 6}$ &  $2$  \\
      $\mathbf{3}_0$ & $0_{\times 3}$ &  $4$  \\
   $\mathbf{5}_0$ & $0_{\times 5}$ &  $4$  \\
  \hline
\end{tabular}
\caption{Scalar masses at the $N=4$ supersymmetric $AdS_5$ critical
point with $U(1)\times SU(2)_{\textrm{diag}}$ symmetry and the
corresponding dimensions of the dual operators.}\label{table2}
\end{table}

\subsection{Supersymmetric RG flows from $N=2$ SCFTs to two dimensional SCFTs}
We now consider another type of solutions namely solutions interpolating between supersymmetric $AdS_5$ vacua identified previously and $AdS_3\times \Sigma_2$ geometries. In the present consideration, $\Sigma_2$ is a two-sphere ($S^2$) or a two-dimensional hyperbolic space ($H^2$). 
\\
\indent We begin with the metric ansatz for the $\Sigma_2=S^2$ case
\begin{equation}
ds^2=e^{2f(r)}dx^2_{1,1}+dr^2+e^{2g(r)}(d\theta^2+\sin^2\theta d\phi^2)\label{metric_ansatz_AdS3}
 \end{equation}
where $dx^2_{1,1}$ is the flat metric in two dimensions. It is useful to note the components of the spin connection
\begin{eqnarray}
\omega^{\hat{\mu}\hat{r}}&=&f'e^{\hat{\mu}},\qquad \omega^{\hat{\theta}\hat{r}}=g'e^{\hat{\theta}},\nonumber \\
\omega^{\hat{\phi}\hat{r}}&=&g'e^{\hat{\phi}},\qquad \omega^{\hat{\phi}\hat{\theta}}=e^{-g}\cot \theta e^{\hat{\phi}}
\end{eqnarray}
with the obvious choice of vielbein
\begin{equation}
e^{\hat{\mu}}=e^fdx^\mu,\qquad e^{\hat{r}}=dr,\qquad e^{\hat{\theta}}=e^gd\theta,\qquad e^{\hat{\phi}}=e^g\sin\theta d\phi
\end{equation}
for $\hat{\mu}=0,1$. 
\\
\indent To preserve some amount of supersymmetry, we impose a twist condition by cancelling the spin connection on $S^2$ with some gauge connections. We will consider abelian twists from $U(1)\times U(1)\times U(1)\subset U(1)\times SU(2)\times SU(2)$ and its $U(1)\times U(1)_{\textrm{diag}}$ subgroup. The corresponding gauge fields are denoted by $(A^0,A^5,A^8)$. Note that turning on $A^0$ and $A^5$ correspond to a twist along the R-symmetry $U(1)\times SU(2)$ of the dual $N=2$ SCFTs while a non-vanishing $A^8$ is related to turning on the gauge field of $SU(2)$ flavor symmetry. The latter cannot be used as a twist since the Killing spinor is neutral under this symmetry.
\\
\indent An effect of the twisting procedure is to cancel $\omega^{\hat{\theta}\hat{\phi}}$ on $S^2$. The BPS conditions $\delta\psi_{i\hat{\theta}}=0$ and $\delta \psi_{i\hat{\phi}}=0$ then lead to the same BPS equation. In order to achieve this, we consider the ansatz for the gauge fields
\begin{equation}
A^{\mc{M}=0,5,8}=a_{\mc{M}}\cos\theta d\phi\, .\label{vector_ansatz}
\end{equation}
We consider two type of solutions with unbroken gauge symmetry $U(1)\times U(1)\times U(1)$ and $U(1)\times U(1)_{\textrm{diag}}$. We begin with a simpler case of $U(1)\times U(1)\times U(1)$ invariant sector consisting of four singlet scalars $\Sigma$ and $\varphi_i$, $i=1,2,3$. The latter correspond to the $SO(5,5)$ non-compact generators $Y_{53}$, $Y_{54}$ and $Y_{55}$. The coset representative is then given by 
\begin{equation}
\mc{V}=e^{\varphi_1 Y_{53}}e^{\varphi_2 Y_{54}}e^{\varphi_3 Y_{55}}\, .\label{U1_3_coset}
\end{equation} 
A straightforward computation gives relevant components of the covariant derivative on the Killing spinors $\epsilon_i$    
\begin{equation}
D_{\hat{\phi}}\epsilon_i= \ldots +\frac{1}{2}e^{-g}\cot\theta\left[\gamma_{\hat{\phi}\hat{\theta}}\epsilon_i-ia_0g_1{(\sigma_2\otimes\sigma_3)_i}^j\epsilon_j+ia_5g_2{(\sigma_1\otimes \sigma_1)_i}^j\epsilon_j\right]. 
\end{equation}
In order to cancel the spin connection, we need to impose the conditions
\begin{eqnarray}
i\gamma_{\hat{\theta}\hat{\phi}}\epsilon_i=a_0g_1{(\sigma_2\otimes\sigma_3)_i}^j\epsilon_j-a_5g_2{(\sigma_1\otimes \sigma_1)_i}^j\epsilon_j\, .\label{twist_con1}
\end{eqnarray}
Consistency with $(i\gamma_{\hat{\theta}\hat{\phi}})^2=\mathbb{I}_4$ requires the conditions
\begin{equation}
(g_1a_0)^2+(g_2a_5)^2=1\qquad \textrm{and}\qquad g_1g_2a_0a_5=0\, .
\end{equation}
The second condition implies, for non-vanishing $g_1$ and $g_2$, either $a_0=0$ or $a_5=0$ for which the first condition gives $g_2a_5=\pm 1$ or $g_1a_0=\pm1$, respectively. These two possibilities correspond respectively to the $\alpha$-twist and $\beta$-twist studied in \cite{alpha_beta_twist}, see also a discussion in \cite{flow_acrossD_bobev}.
\\
\indent For $a_0=0$ and $g_2a_5=\pm 1$, the condition \eqref{twist_con1} becomes a projector
\begin{equation}
i\gamma_{\hat{\theta}\hat{\phi}}\epsilon_i=\mp {(\sigma_1\otimes \sigma_1)_i}^j\epsilon_j\, .\label{gamma_theta_phi_projector}
\end{equation}
For $a_5=0$ and $g_1a_0=\pm 1$, we find
\begin{equation}
i\gamma_{\hat{\theta}\hat{\phi}}\epsilon_i=\pm{(\sigma_2\otimes \sigma_3)_i}^j\epsilon_j\, .
\end{equation}
It should be noted that we can make a definite sign choice for the twist condition and the $\gamma_{\hat{\theta}\hat{\phi}}$ projector. The other possiblility can be obtained by changing the sign of $a_0$ or $a_5$ together with a sign change in the $\gamma_{\hat{\theta}\hat{\phi}}$ projector. In the remaining part of this paper, we will choose the twist conditions and $\gamma_{\hat{\theta}\hat{\phi}}$ projector with the upper sign.
\\
\indent For the $U(1)\times U(1)_{\textrm{diag}}$ sector with the $U(1)_{\textrm{diag}}$ being a diagonal subgroup of $U(1)\times U(1)\subset SU(2)\times SU(2)$, we have five singlets from the vector multiplet scalars corresponding to the following non-compact generators of $SO(5,5)$
\begin{eqnarray}
\hat{Y}_1&=&Y_{31}+Y_{42},\qquad \hat{Y}_2=Y_{53},\qquad \hat{Y}_3=Y_{32}-Y_{41},\nonumber \\
\hat{Y}_4&=&Y_{54},\qquad \hat{Y}_5=Y_{55}
\end{eqnarray}
giving rise to the coset representative
\begin{equation}
\mc{V}=e^{\phi_1\hat{Y}_1}e^{\phi_2\hat{Y}_2}e^{\phi_3\hat{Y}_3}e^{\phi_4\hat{Y}_4}e^{\phi_5\hat{Y}_5}\, .\label{U1_U1d_coset}
\end{equation}
The result of the analysis is the same as in the previous case but with an additional condition imposed on the flux parameters $a_5$ and $a_8$
\begin{equation}
g_2a_5=g_3a_8
\end{equation}
implementing the $U(1)_{\textrm{diag}}$ gauge symmetry. It turns out that, in both cases, all two-form fields can be consistently set to zero provided that $A^1$ and $A^2$ vanish.  
\\
\indent For the $H^2$ case, we simply change $\sin\theta$ to $\sinh\theta$ in the metric \eqref{metric_ansatz_AdS3} and take the gauge fields to be $A^{\mc{M}}=a_{\mc{M}}\cosh\theta d\phi$. The twist procedure works as in the $S^2$ case. However, due to the opposite sign in the covariant field strengths $\mc{H}^{\mc{M}}=dA^{\mc{M}}$, the resulting BPS equations for the two cases are related to each other by a sign change in the twist parameters $a_{\mc{M}}$. 

\subsubsection{Flow solutions with $U(1)\times U(1)\times U(1)$ symmetry}
\begin{figure}
         \centering
         \begin{subfigure}[b]{0.45\textwidth}
                 \includegraphics[width=\textwidth]{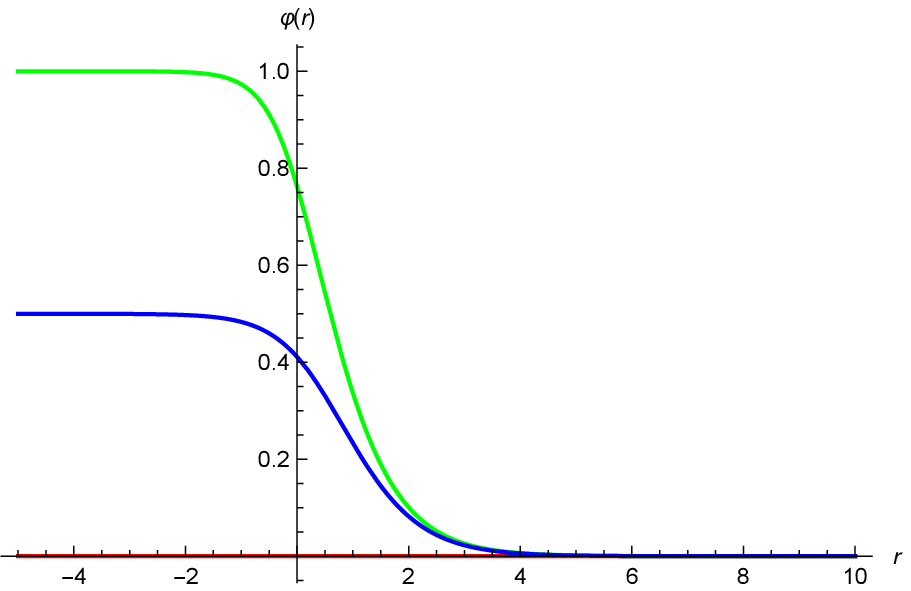}
                 \caption{Solution for $\varphi$}
         \end{subfigure} \qquad
\begin{subfigure}[b]{0.45\textwidth}
                 \includegraphics[width=\textwidth]{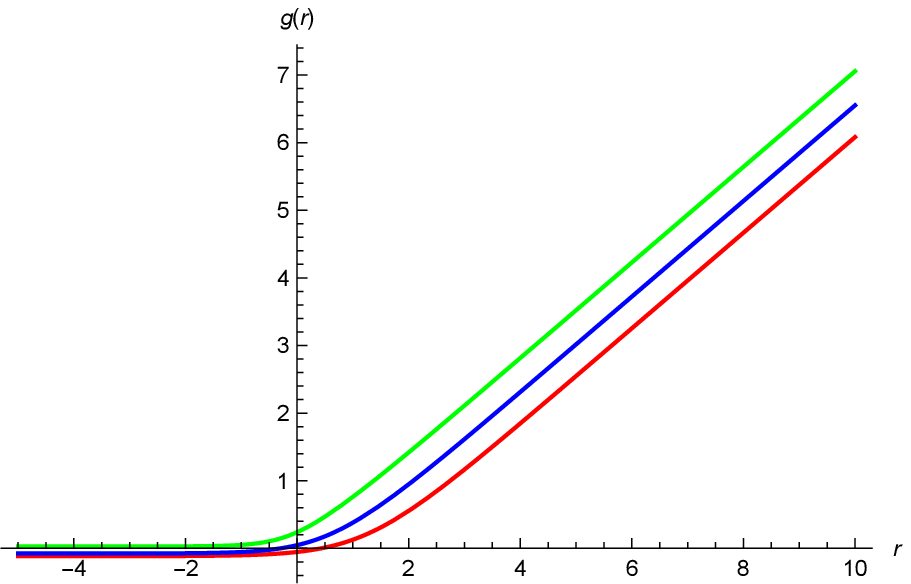}
                 \caption{Solution for $g$}
         \end{subfigure}

         ~ %add desired spacing between images, e. g. ~, \quad, \qquad, \hfill etc.
         \begin{subfigure}[b]{0.45\textwidth}
                 \includegraphics[width=\textwidth]{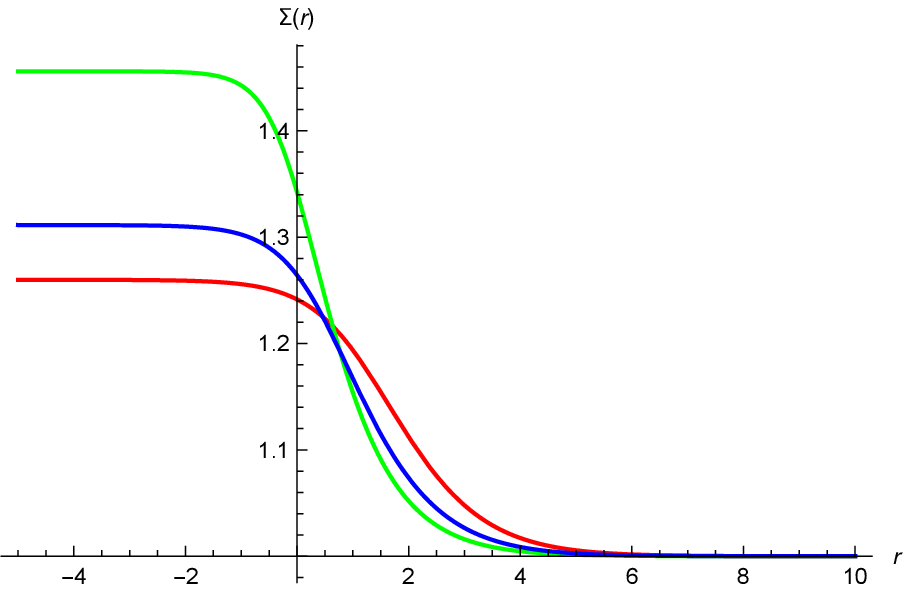}
                 \caption{Solution for $\Sigma$}
         \end{subfigure}\qquad 
         \begin{subfigure}[b]{0.45\textwidth}
                 \includegraphics[width=\textwidth]{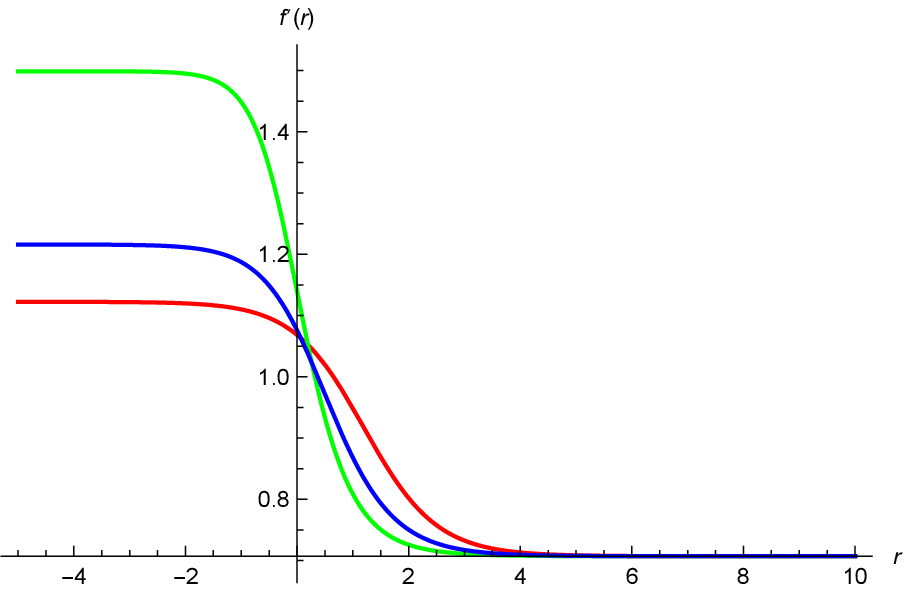}
                 \caption{Solution for $f'$}
         \end{subfigure}
         \caption{RG flows from $N=4$ $AdS_5$ critical point with $U(1)\times SU(2)\times SU(2)$ symmetry to $N=4$ $AdS_3\times H^2$ geometries in the IR with $g_1=1$ and $\varphi_0=0$ (red), $\varphi_0=0.5$ (blue) and $\varphi_0=1$ (green).}\label{fig1}
 \end{figure}

With the coset representative \eqref{U1_3_coset}, the scalar potential and the superpotential are given respectively by
\begin{equation}
V=-\frac{1}{2}\Sigma^{-2}(g_2^2-2\sqrt{2}g_1g_2\Sigma^3\cosh \varphi_1\cosh \varphi_2\cosh \varphi_3)
\end{equation}
and
\begin{equation}
W=\frac{1}{6}\Sigma^{-1}(\sqrt{2}g_1\Sigma^3-2g_2\cosh\varphi_1\cosh \varphi_2\cosh \varphi_3).
\end{equation}
The scalar kinetic term reads
\begin{equation}
\mc{L}_{\textrm{kin}}=-\frac{3}{2}\Sigma^{-2}\Sigma'^2-\frac{1}{2}\cosh^2\varphi_2\cosh^2\varphi_3\varphi_1'^2-\frac{1}{2}\cosh^2\varphi_3\varphi_2'^2-\frac{1}{2}\varphi_3'^2\, .
\end{equation}
The scalar potential can also be written in term of the superpotential as
\begin{eqnarray}
V&=&\frac{3}{2}\Sigma^2\left(\frac{\pd W}{\pd \Sigma}\right)^2+\frac{9}{2}\cosh^{-2}\varphi_2\cosh^{-2}\varphi_3\left(\frac{\pd W}{\pd\varphi_1}\right)^2\nonumber \\
& &+\frac{9}{2}\cosh^{-2}\varphi_3\left(\frac{\pd W}{\pd\varphi_2}\right)^2+\frac{9}{2}\left(\frac{\pd W}{\pd\varphi_3}\right)^2-6W^2\, .
\end{eqnarray}
It can be easily checked that setting $\varphi_2=\varphi_3=0$ is a consistent truncation. Moreover, the result with non-vanishing $\varphi_2$ and $\varphi_3$ is not significantly different from that with $\varphi_2=\varphi_3=0$. Therefore, we will further simplify the computation by using this truncation and set $\varphi_1=\varphi$.
\\
\indent We first consider the case with $a_0=0$. By using the $\gamma_{\hat{r}}$ projection \eqref{gamma_r_projector} with the matrix ${I_i}^j$ given in \eqref{I_gamma_r}, we find the following BPS equations
\begin{eqnarray}
\varphi'&=&\frac{1}{2}\Sigma^{-1}e^{-\varphi-2g}[g_2e^{2g}(e^{2\varphi}-1)-\kappa\Sigma^2(a_5-a_8)+\kappa(a_5+a_8)e^{2\varphi}],\\
\Sigma'&=&-\frac{1}{3}\left[\sqrt{2}g_1\Sigma^3+g_2\cosh\varphi-\kappa e^{-2g}\Sigma^2(a_5\cosh\varphi+a_8\sinh\varphi)\right],\\
g'&=&\frac{1}{6}\Sigma^{-1}\left[\sqrt{2}g_1\Sigma^3-2g_2\cosh\varphi-\kappa 4e^{-2g}\Sigma^2(a_5\cosh\varphi+a_8\sinh\varphi)\right],\\
f'&=&\frac{1}{6}\Sigma^{-1}\left[\sqrt{2}g_1\Sigma^3+2\kappa g_2\cosh\varphi+2\kappa e^{-2g}\Sigma^2(a_5\cosh\varphi+a_8\sinh\varphi)\right].
\end{eqnarray}
The sign choices $\kappa=+1$ and $\kappa=-1$ correspond to $\Sigma_2=S^2$ and $\Sigma_2=H^2$, respectively. We will use this convention throughout the paper.
\\
\indent The $AdS_3\times \Sigma_2$ vacua are characterized by the conditions $g'=\varphi'=\Sigma'=0$ and $f'=\frac{1}{L_{AdS_3}}$. It turns out that the above equations admit any $AdS_3$ solutions only for $a_8=0$ and $\kappa=-1$. In this case, we find that any constant value of $\varphi$ leads to an $AdS_3\times H^2$ solution of the form 
\begin{eqnarray}
\varphi&=&\varphi_0,\qquad \Sigma=-\left(\frac{\sqrt{2}g_2\cosh\varphi_0}{g_1}\right)^{\frac{1}{3}},\nonumber \\
g&=&\frac{1}{6}\ln\left[\frac{2a_5\cosh^2\varphi_0}{g_1^2g_2}\right],\qquad L_{AdS_3}=\left(\frac{\sqrt{2}}{g_1g_2^2\cosh^2\varphi_0}\right)^{\frac{1}{3}}
\end{eqnarray}   
where $\varphi_0$ is a constant. This solution preserves eight supercharges or $N=4$ in three dimensions due to the $\gamma_{\hat{\theta}\hat{\phi}}$ projector. On the other hand, the entire flow solution will preserve only four supercharges due to an additional $\gamma_{\hat{r}}$ projector. 
\\
\indent For $\varphi_0=0$, the solution has $U(1)\times U(1)\times SU(2)$ symmetry due to the vanishing $A^8$ while the solution with $\varphi_0\neq 0$ has smaller symmetry $U(1)\times U(1)\times U(1)$. The resulting $AdS_3\times H^2$ geometry should be dual to a two dimensional $N=(2,2)$ SCFT with $SU(2)$ or $U(1)$ flavor symmetry depending on the value of $\varphi_0$. An asymptotic analysis near the $AdS_3\times H^2$ critical point shows that $\varphi$ is dual to a marginal operator in the two-dimensional SCFT. The central charge of the dual SCFT can also be computed by \cite{2D_central_charge}
\begin{equation}
 c=\frac{3L_{AdS_3}}{2G_3}= \frac{3L_{AdS_3}\textrm{vol}(H^2)}{2G_5}e^{2g_0}=\frac{3\textrm{vol}(H^2) a_5^{\frac{1}{3}}}{\sqrt{2}g_1g_2G_5}
\end{equation}
which is independent of $\varphi_0$. $g_0$ is the value of $g(r)$ at the $AdS_3$ critical point. For $H_2$ being a genus $\mathfrak{g}>1$ Riemann surface, we have $\textrm{vol}(H^2)=4\pi (\mathfrak{g}-1)$. 
\\
\indent Examples of numerical flow solutions interpolating between $N=4$ supersymmetric $AdS_5$ and $N=4$ supersymmetric $AdS_3\times H^2$ with different values of $\varphi_0$ are given in figure \ref{fig1}. The solution with $\varphi_0=0$ is effectively the same as that studied in \cite{flow_acrossD_bobev} which is in turn obtained from the solutions in \cite{5D_N4_Romans} by turning off the $U(1)$ gauge field. In this case, the matter multiplets can be decoupled. Solutions with $\varphi_0\neq 0$ are only possible in the matter-coupled gauged supergravity and have not previously appeared.  
\\
\indent For the case of $a_5=0$, we also find that the BPS conditions require $a_8=0$. The resulting BPS equations read
\begin{eqnarray}
\varphi'&=&g_2\Sigma^{-1}\sinh\varphi,\\
\Sigma'&=&-\frac{\sqrt{2}}{3}\Sigma^{-1}(\kappa a_0e^{-2g}+g_1\Sigma^4)-\frac{1}{3}g_2\cosh\varphi,\\
g'&=&\frac{1}{6}\Sigma^{-2}(\sqrt{2}g_1\Sigma^4-2g_2\Sigma\cosh\varphi-2\kappa\sqrt{2}a_0e^{-2g}),\\
f'&=&\frac{1}{6}\Sigma^{-2}(\sqrt{2}g_1\Sigma^4-2g_2\Sigma\cosh\varphi+\sqrt{2}\kappa a_0e^{-2g}).
\end{eqnarray}       
Note also that the BPS equation for $\varphi$ does not involve $a_0$ since $\varphi$ is neutral under $A^0$. In this case, $AdS_3$ vacua do not exist. For $\varphi'=g'=0$, we find a singular behavior of $\Sigma$ at a finite value of $r$
\begin{equation}
\Sigma\sim \frac{1}{\sqrt{\sqrt{2}g_1r-C}}
\end{equation}
for some contant $C$. This has also been pointed out in \cite{flow_acrossD_bobev}. 

\subsubsection{Flow solutions with $U(1)\times U(1)_{\textrm{diag}}$ symmetry}
\begin{figure}
         \centering
         \begin{subfigure}[b]{0.3\textwidth}
                 \includegraphics[width=\textwidth]{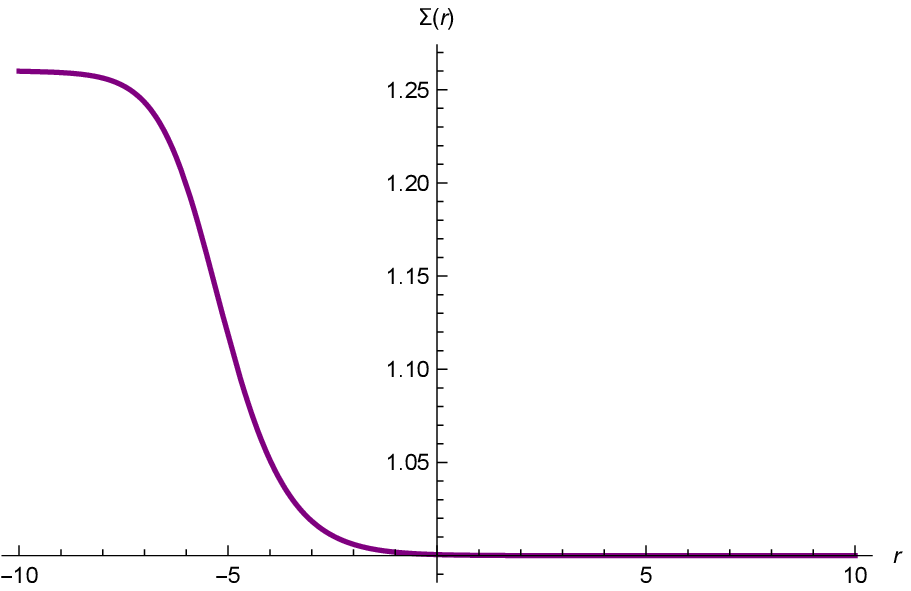}
                 \caption{Solution for $\Sigma$}
         \end{subfigure} \quad
\begin{subfigure}[b]{0.3\textwidth}
                 \includegraphics[width=\textwidth]{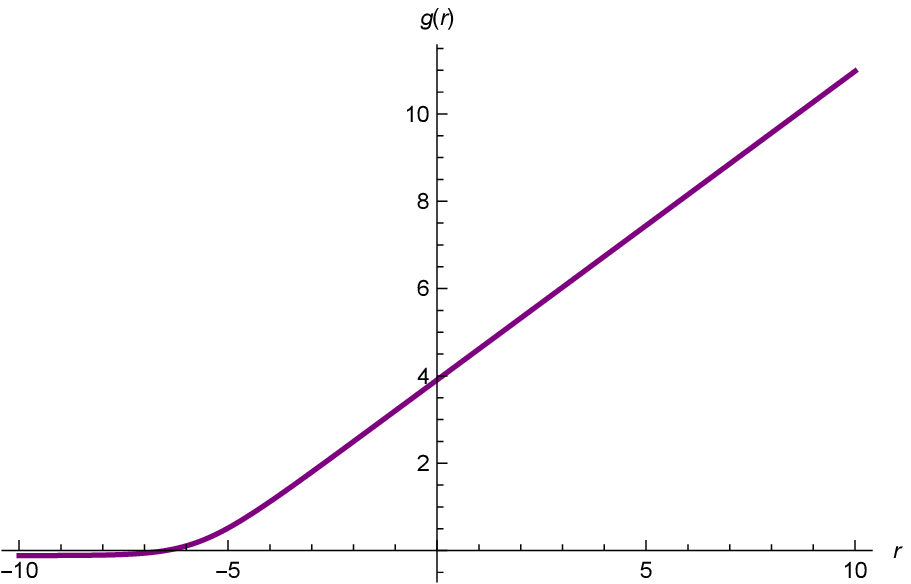}
                 \caption{Solution for $g$}
         \end{subfigure}\quad
         \begin{subfigure}[b]{0.3\textwidth}
                 \includegraphics[width=\textwidth]{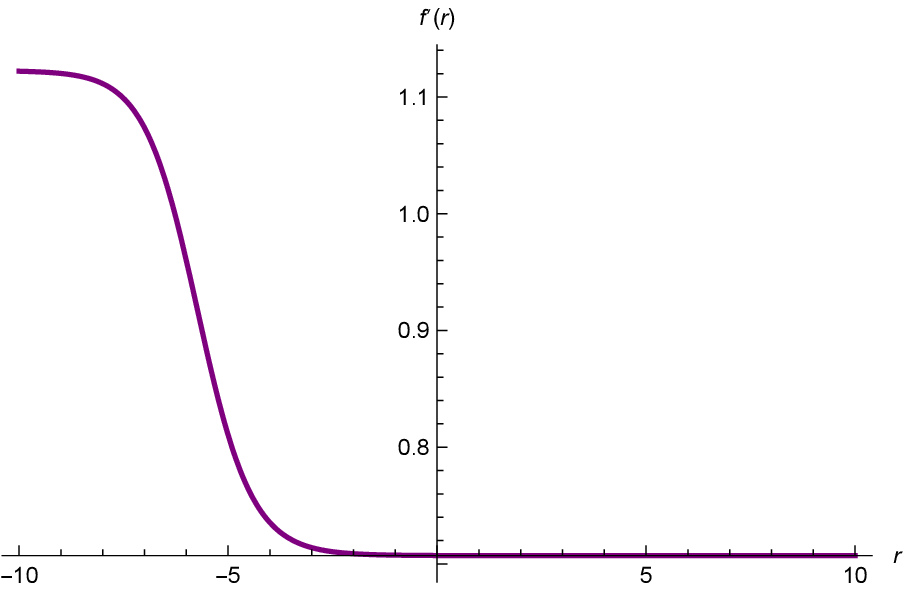}
                 \caption{Solution for $f'$}
         \end{subfigure}\qquad 
         \caption{An RG flow between $N=4$ $AdS_5$ vacuum with $U(1)\times SU(2)\times SU(2)$ symmetry and $N=4$ $AdS_3\times H^2$ vacuum with $U(1)\times U(1)_{\textrm{diag}}$ symmetry in \eqref{AdS3_compact1} for $g_1=1$.}\label{fig2}
 \end{figure}

\begin{figure}
         \centering
         \begin{subfigure}[b]{0.3\textwidth}
                 \includegraphics[width=\textwidth]{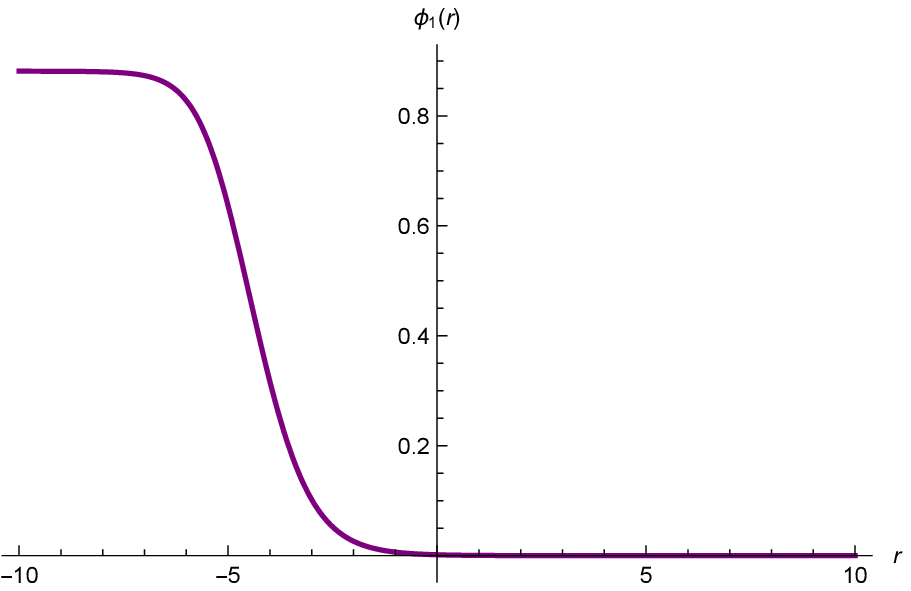}
                 \caption{Solution for $\phi_1$}
         \end{subfigure} \,\,\,
\begin{subfigure}[b]{0.3\textwidth}
                 \includegraphics[width=\textwidth]{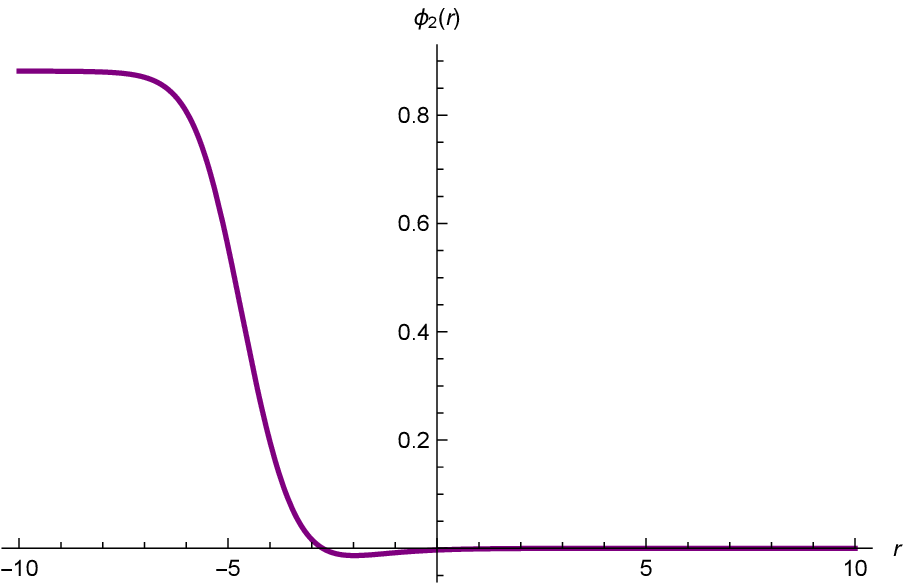}
                 \caption{Solution for $\phi_2$}
         \end{subfigure}\,\,\,
\begin{subfigure}[b]{0.3\textwidth}
                 \includegraphics[width=\textwidth]{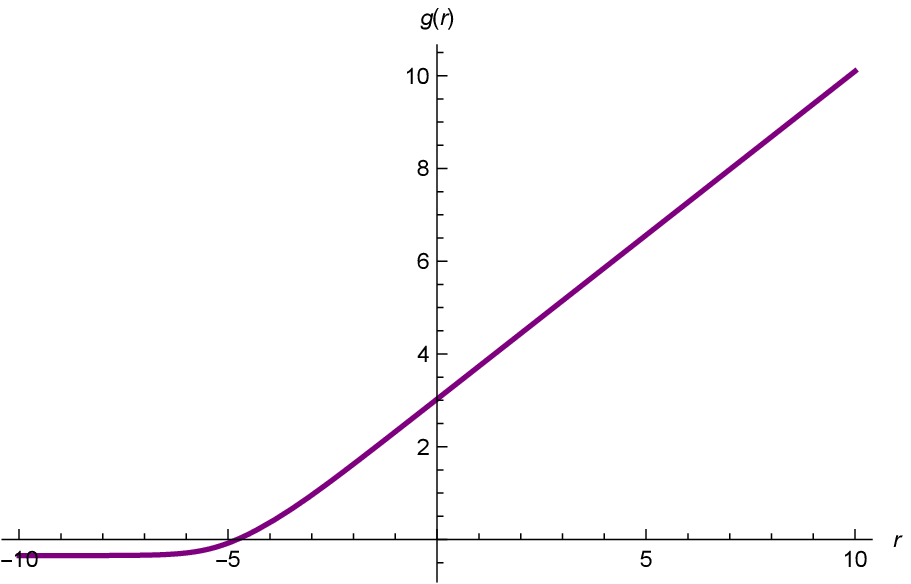}
                 \caption{Solution for $g$}
         \end{subfigure}\\

         ~ %add desired spacing between images, e. g. ~, \quad, \qquad, \hfill etc.
         \begin{subfigure}[b]{0.45\textwidth}
                 \includegraphics[width=\textwidth]{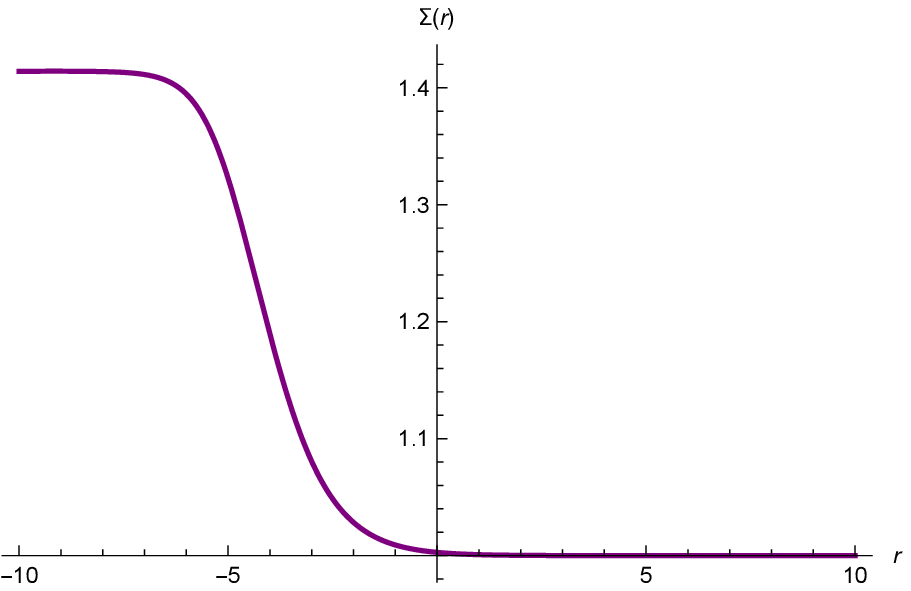}
                 \caption{Solution for $\Sigma$}
         \end{subfigure}\qquad 
         \begin{subfigure}[b]{0.45\textwidth}
                 \includegraphics[width=\textwidth]{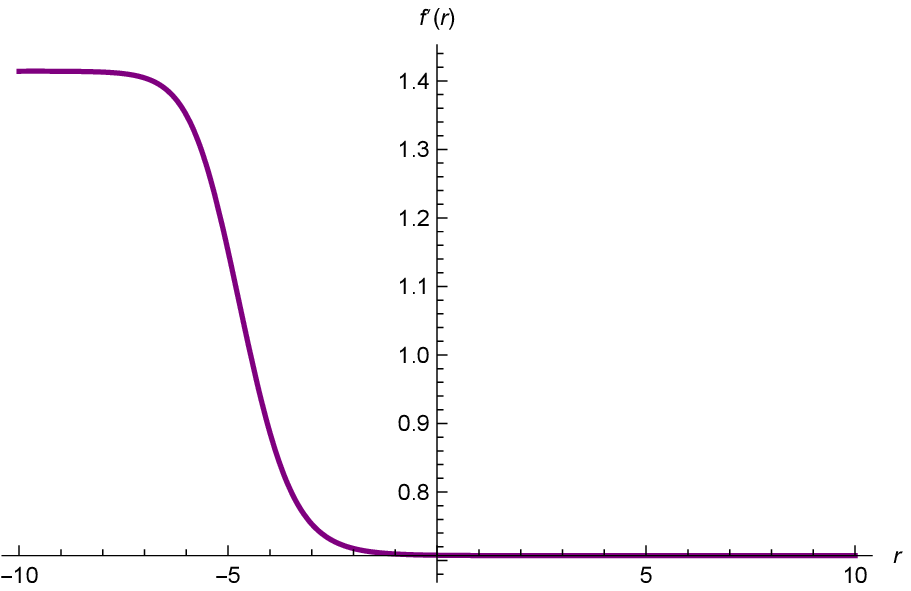}
                 \caption{Solution for $f'$}
         \end{subfigure}
         \caption{An RG flow from $AdS_5$ critical point with $U(1)\times SU(2)\times SU(2)$ symmetry to $AdS_3\times H^2$ critical point II for $g_1=1$ and $g_3=2g_1$.}\label{fig3}
 \end{figure}

\begin{figure}
         \centering
         \begin{subfigure}[b]{0.3\textwidth}
                 \includegraphics[width=\textwidth]{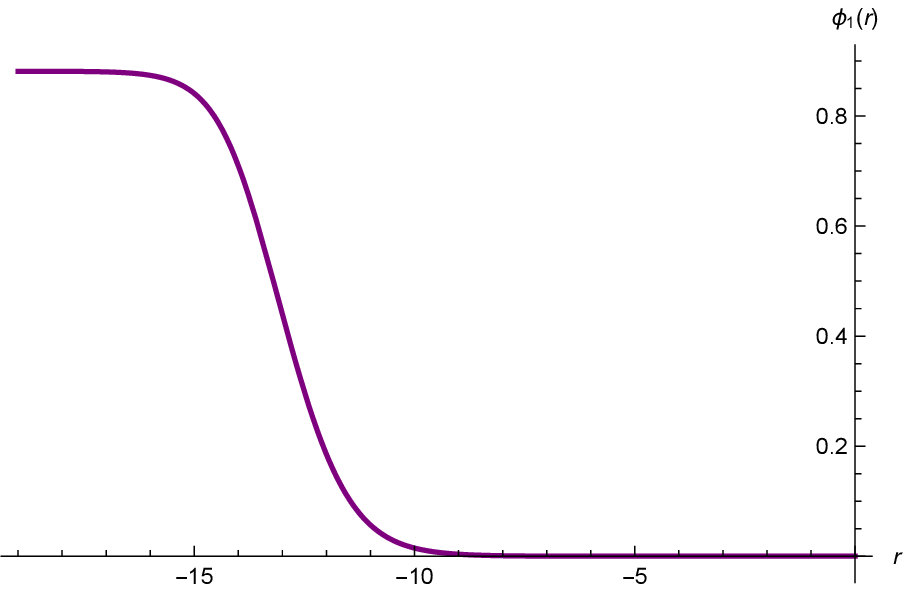}
                 \caption{Solution for $\phi_1$}
         \end{subfigure} \,\,\,
\begin{subfigure}[b]{0.3\textwidth}
                 \includegraphics[width=\textwidth]{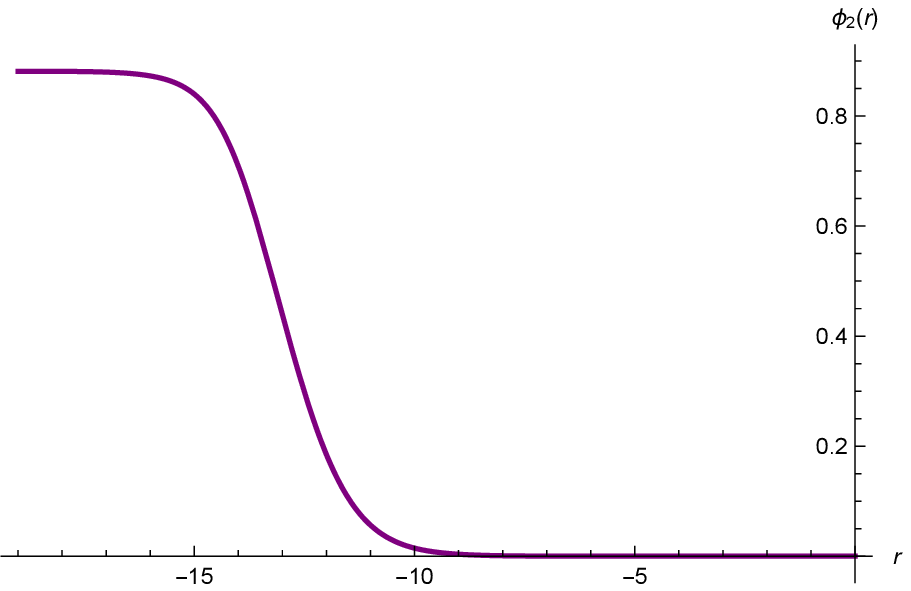}
                 \caption{Solution for $\phi_2$}
         \end{subfigure}\,\,\,
\begin{subfigure}[b]{0.3\textwidth}
                 \includegraphics[width=\textwidth]{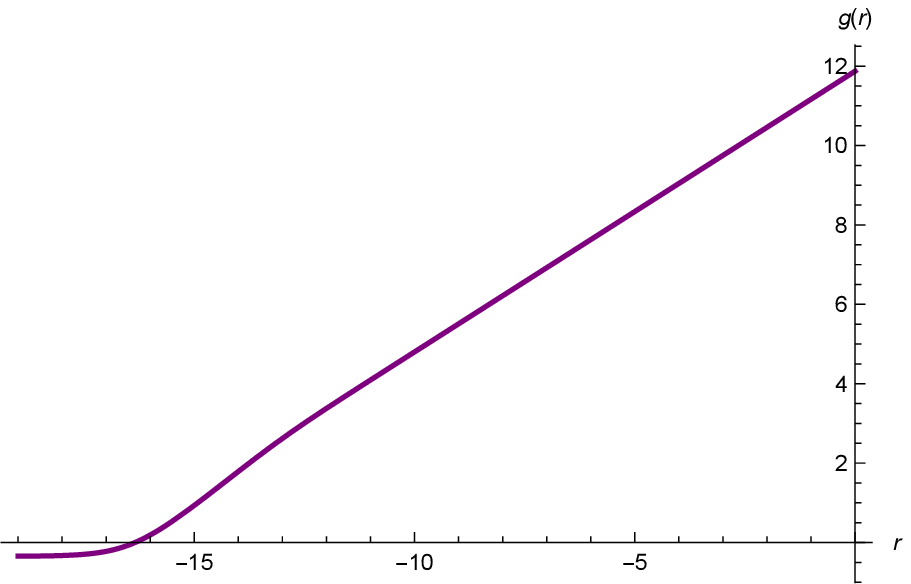}
                 \caption{Solution for $g$}
         \end{subfigure}\\

         ~ %add desired spacing between images, e. g. ~, \quad, \qquad, \hfill etc.
         \begin{subfigure}[b]{0.45\textwidth}
                 \includegraphics[width=\textwidth]{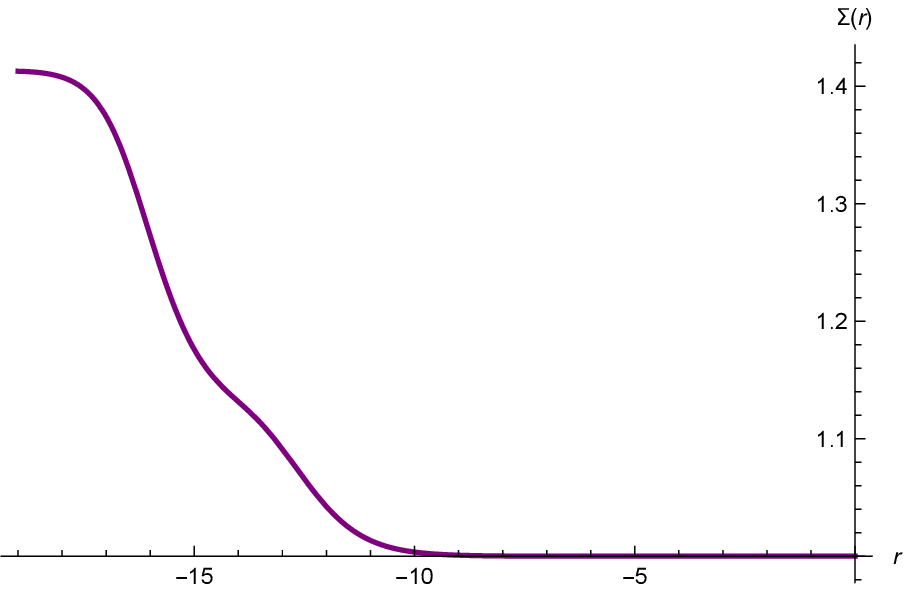}
                 \caption{Solution for $\Sigma$}
         \end{subfigure}\qquad 
         \begin{subfigure}[b]{0.45\textwidth}
                 \includegraphics[width=\textwidth]{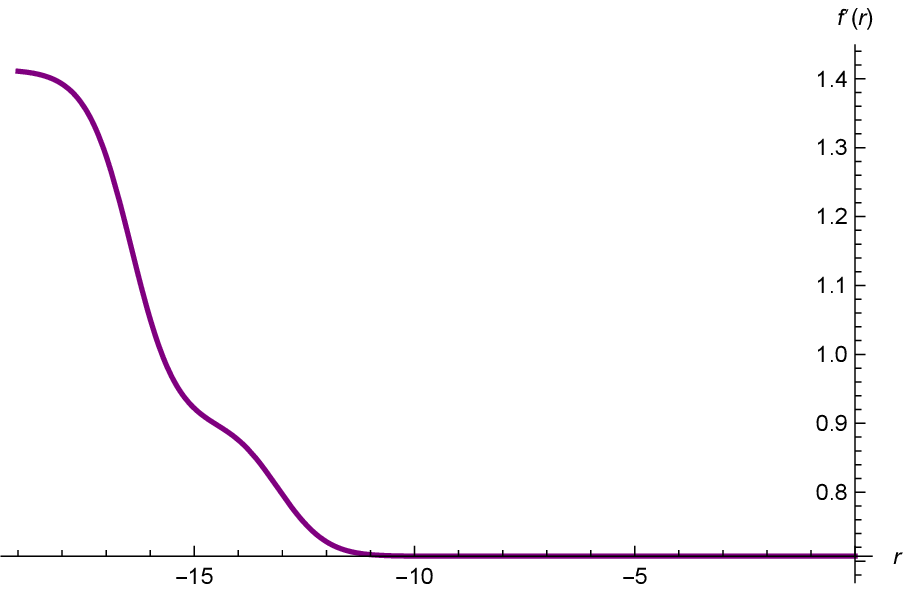}
                 \caption{Solution for $f'$}
         \end{subfigure}
         \caption{An RG flow from $AdS_5$ critical point with $U(1)\times SU(2)\times SU(2)$ symmetry to $AdS_5$ critical point with $U(1)\times SU(2)_{\textrm{diag}}$ symmetry and finally to $AdS_3\times H^2$ critical point II for $g_1=1$ and $g_3=2g_1$.}\label{fig4}
 \end{figure}

In this case, there are five singlet scalars from $SO(5,5)/SO(5)\times SO(5)$ coset with the coset representative given by \eqref{U1_U1d_coset}. Together with $\Sigma$, there are in total six singlet scalars, and the computation is much more complicated than the previous case. We will again make a truncation by setting $\phi_4=\phi_5=0$ in the following analysis. The scalar potential with this truncation is given by
\begin{eqnarray}
V&=&\frac{1}{16}\Sigma^{-2}\left[8\sqrt{2}g_1(g_2\cosh\phi_2-g_3\sinh\phi_2)+4(g_3^2-g_2^2) \right.\nonumber \\
& &+g_2g_3\sinh(2\phi_2)(2\cosh^2(2\phi_1)\cosh(4\phi_3)+\cosh(4\phi_1)-3)\nonumber \\
& &+8\sqrt{2}g_1\cosh(2\phi_1)\cosh(2\phi_3)\Sigma^3(g_2\cosh\phi_2+g_3\sinh\phi_2)\nonumber \\
& &+(g_2^2+g_3^2)\cosh(2\phi_2)(\cosh^2(2\phi_1)\cosh^2(2\phi_3)-1)\nonumber \\
& &\left.-4(g_2^2+g_3^2)\cosh(2\phi_1)\cosh(2\phi_3)
\right].
\end{eqnarray}
This can be written in term of the superpotential as
\begin{equation}
V=\frac{3}{2}\Sigma^2\left(\frac{\pd W}{\pd \Sigma}\right)^2+\frac{9}{4}\cosh^{-2}(2\phi_3)\left(\frac{\pd W}{\pd \phi_1}\right)^2+\frac{9}{2}\left(\frac{\pd W}{\pd \phi_2}\right)^2+\frac{9}{4}\left(\frac{\pd W}{\pd \phi_3}\right)^2-6W^2
\end{equation}
where the superpotential in this case is given by
\begin{eqnarray} 
W&=&\frac{1}{6}\Sigma^{-1}\left[\sqrt{2}g_1\Sigma^3-g_2\cosh\phi_2+g_3\sinh\phi_2 \right.\nonumber \\
& &\left.\phantom{\sqrt{2}} -\cosh(2\phi_1)\cosh(2\phi_3)(g_2\cosh\phi_2+g_3\sinh\phi_2)\right].
\end{eqnarray}
It can be verified that this superpotential admits two critical points given in equations \eqref{AdS5_1_compact} and \eqref{AdS5_2_compact}. When $\phi_1=\phi_3=0$, this is the $U(1)\times U(1)\times U(1)$ invariant sector. For $\phi_3=0$ and $\phi_1=\phi_2$, we reobtain the $U(1)\times SU(2)_{\textrm{diag}}$ invariant scalars which admit the second $N=4$ $AdS_5$ critical point with $U(1)\times SU(2)_{\textrm{diag}}$ symmetry.
\\
\indent We firstly consider the twist from $A^0$ gauge field. For $a_5=0$, the $U(1)_{\textrm{diag}}$ symmetry also demands $a_8=0$. The BPS equations for $\phi_1$, $\phi_2$ and $\phi_3$ will not depend on the twist parameter $a_0$ since they are not charged under $A^0$. Therefore, the only possibility to have $AdS_3$ vacua is to set these scalars to their values at the two $AdS_5$ critical points. Setting all $\phi_i=0$ for $i=1,2,3$ dose not lead to any $AdS_3$ solutions as in the previous case. The other choice namely $\phi_3=0$ and $\phi_1=\phi_2=\frac{1}{2}\ln\left[\frac{g_3-g_2}{g_3+g_2}\right]$ does not give rise to any $AdS_3$ vacua either. Therefore, we will not give the explicit form of the BPS equations in this case.     
\\
\indent We now consider the twist from $A^5$ and $A^8$ gauge fields. In this case, we do find some $AdS_3$ solutions. The BPS equations read
\begin{eqnarray}
\phi_1'&=&\frac{1}{2}\Sigma^{-1}\textrm{sech}(2\phi_3)\sinh(2\phi_1)(g_2\cosh\phi_2+g_3\sinh\phi_2),\\
\phi_2'&=&\frac{1}{2}\Sigma^{-1}\left[\cosh(2\phi_1)\cosh(2\phi_3)(g_3\cosh\phi_2+g_2\sinh\phi_2)-g_3\sinh\phi_2 \right. \nonumber \\
& &\left.-2\kappa\Sigma^2e^{-2g}(a_8\cosh\phi_2+a_5\sinh\phi_2) +g_2\sinh\phi_2\right],\\
\phi_3'&=&\frac{1}{2}\Sigma^{-1}\cosh(2\phi_1)\sinh(2\phi_3)(g_2\cosh\phi_2+g_3\sinh\phi_2),\\
\Sigma'&=&\frac{1}{6}\left[g_3\sinh\phi_2 -\cosh(2\phi_2)\cosh(2\phi_3)(g_2\cosh\phi_2+g_3\sinh\phi_2)\right. \nonumber \\
& &\left.-2\kappa e^{-2g}\Sigma^2(a_5\cosh\phi_2+a_8\sinh\phi_2)-2\sqrt{2}\Sigma^3-g_2\cosh\phi_2\right],\\
g'&=&\frac{1}{6}\Sigma^{-1}\left[g_3\sinh\phi_2-\cosh(2\phi_1)\cosh(2\phi_3)(g_2\cosh\phi_2+g_3\sinh\phi_2) \right. \nonumber \\
& &\left.-g_2\cosh\phi_2+4\kappa e^{-2g}\Sigma^2(a_5\cosh\phi_2+a_8\sinh\phi_2)+\sqrt{2}g_1\Sigma^3\right],\\
f'&=&\frac{1}{6}\Sigma^{-1}\left[g_3\sinh\phi_2-\cosh(2\phi_1)\cosh(2\phi_3)(g_2\cosh\phi_2+g_3\sinh\phi_2) \right. \nonumber \\
& &\left.-g_2\cosh\phi_2-2\kappa e^{-2g}\Sigma^2(a_5\cosh\phi_2+a_8\sinh\phi_2)+\sqrt{2}g_1\Sigma^3\right]
\end{eqnarray}
for which there is a relation $g_2a_5=g_3a_8$ to be imposed.
\\
\indent We find that these equations admit $AdS_3\times \Sigma_2$ solutions only for $\kappa=-1$. The $AdS_3\times H^2$ solutions are given as follow.
\begin{itemize}
\item I. The simplest solution is obtained by setting $\phi_i=0$, $i=1,2,3$ and
\begin{equation}
\Sigma=-\left(\frac{\sqrt{2}g_2}{g_1}\right)^{\frac{1}{3}},\quad g=\frac{1}{6}\ln\left[\frac{2a_5}{g_1^2g_2}\right],\quad L_{AdS_3}=\left(\frac{\sqrt{2}}{g_1g_2^2}\right)^{\frac{1}{3}}\, .\label{AdS3_compact1}
\end{equation}
\item II. One of the $AdS_3\times H^2$ solutions with vector multiplet scalars non-vanishing is given by
\begin{eqnarray}
\phi_1&=&\phi_2=\frac{1}{2}\ln\left[\frac{g_3-g_2}{g_3+g_2}\right],\qquad \phi_3=0,\qquad \Sigma^3=-\frac{\sqrt{2}g_2g_3}{g_1\sqrt{g_3^2-g_2^2}},\nonumber \\
g&=&\frac{1}{6}\ln\left[\frac{a_5^3(g_3^2-g_2^2)^2}{g_1^2g_2g_3^4}\right],\qquad L_{AdS_3}=\left(\frac{\sqrt{2}(g_3^2-g_2^2)}{g_1g_2^2g_3^2}\right)^{\frac{1}{3}}\, . \label{AdS3_compact2}
\end{eqnarray}
\item III. There is another $AdS_3\times H^2$ solution located at
\begin{eqnarray}
\phi_1&=&0,\quad \phi_2=\phi_3=\frac{1}{2}\ln\left[\frac{g_3-g_2}{g_3+g_2}\right],\quad \Sigma^3=-\frac{\sqrt{2}g_2g_3}{g_1\sqrt{g_3^2-g_2^2}},\nonumber \\
g&=&\frac{1}{6}\ln\left[\frac{2a_5^3(g_3^2-g_2^2)^2}{g_1^2g_2g_3^4}\right],\qquad L_{AdS_3}=\left(\frac{\sqrt{2}(g_3^2-g_2^2)}{g_1g_2^2g_3^2}\right)^{\frac{1}{3}}\, .\label{AdS3_compact3}
\end{eqnarray}
\end{itemize}  
All of these solutions preserve eight supercharges corresponding to $N=4$ supersymmetry in three dimensions or equivalently $N=(2,2)$ in the dual two dimensional SCFTs. It should also be noted that critical points II and III appear to be related by a permutation of $\phi_i$. However, the solution with $\phi_2=0$ does not exist since this also requires $\phi_1=\phi_3=0$ and $a_8=a_5=0$.
\\
\indent The next step is to find RG flow solutions interpolating between $N=4$ supersymmetric $AdS_5$ critical points and the above $AdS_3\times H^2$ geometries. We first consider a simple case of the flow to $AdS_3\times H^2$ critical point I with $\phi_1=\phi_2=\phi_3=0$. The BPS equations simplify considerably to
\begin{eqnarray}
\Sigma'&=&-\frac{1}{3}(\sqrt{2}g_1\Sigma^3+g_2-a_5e^{-2g}\Sigma^2),\\
g'&=&\frac{1}{6}\Sigma^{-1}(\sqrt{2}g_1\Sigma^3-2g_2-4a_5e^{-2g}\Sigma^2),\\
f'&=&\frac{1}{6}\Sigma^{-1}(\sqrt{2}g_1\Sigma^3-2g_2+2a_5e^{-2g}\Sigma^2).
\end{eqnarray}
We can partially solve these equations analytically and find a relation between solutions of $g$ and $\Sigma$ of the form
\begin{equation}
2a_5g+4a_5\ln\Sigma=e^{2g}\Sigma^{-2}(g_2+\sqrt{2}g_1\Sigma^3).
\end{equation} 
However, the complete solutions can only be found numerically. In this case, the solutions reduce to the universal flows across dimensions considered in \cite{flow_acrossD_bobev}. An example of these solutions is given in figure \ref{fig2}.    
\\
\indent $AdS_3\times H^2$ critical point II is more interesting in the sense that it can be connected to both of the $N=4$ $AdS_5$ vacua. In order to obtain RG flow solutions, we set $\phi_3=0$ which is a consistent truncation. An example of flows from $AdS_5$ with $U(1)\times SU(2)\times SU(2)$ symmetry to $AdS_3\times H^2$ critical point II is given in figure \ref{fig3}. With suitable boundary conditions, we can find a solution that flows from $AdS_5$ with $U(1)\times SU(2)\times SU(2)$ symmetry and approaches $AdS_5$ with $U(1)\times SU(2)_{\textrm{diag}}$ symmetry before reaching the $AdS_3\times H^2$ critical point II. A solution of this type is shown in figure \ref{fig4}.      

\begin{figure}
         \centering
         \begin{subfigure}[b]{0.3\textwidth}
                 \includegraphics[width=\textwidth]{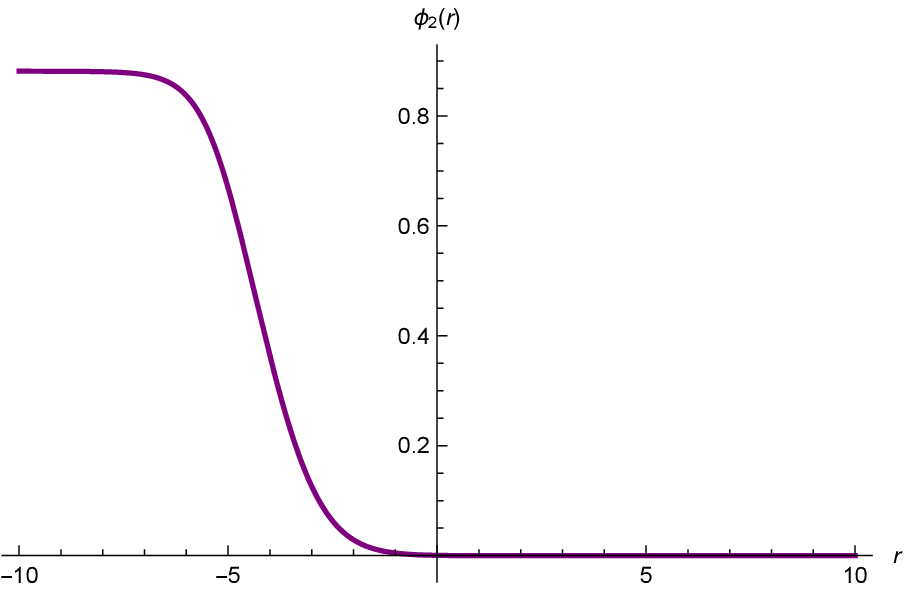}
                 \caption{Solution for $\phi_2$}
         \end{subfigure} \,\,\,
\begin{subfigure}[b]{0.3\textwidth}
                 \includegraphics[width=\textwidth]{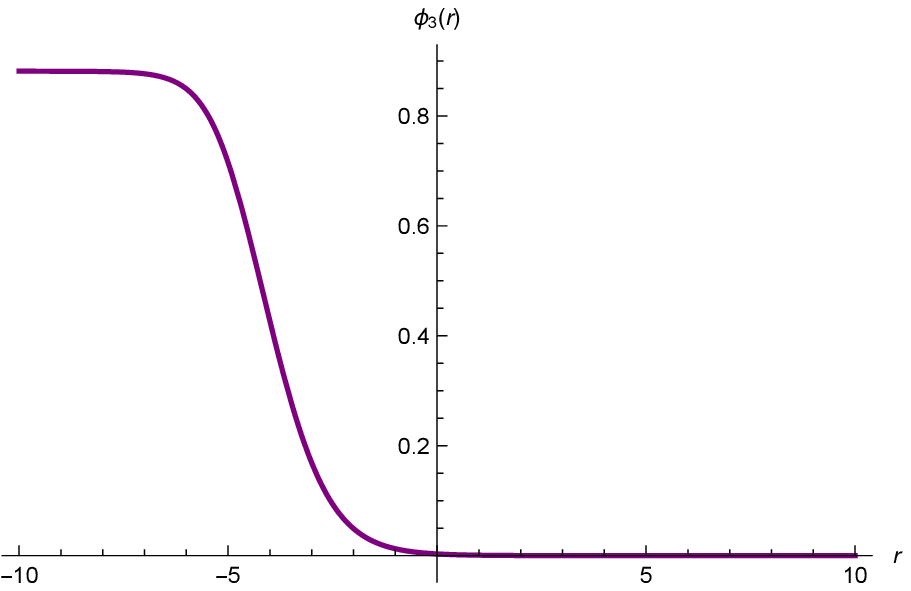}
                 \caption{Solution for $\phi_3$}
         \end{subfigure}\,\,\,
\begin{subfigure}[b]{0.3\textwidth}
                 \includegraphics[width=\textwidth]{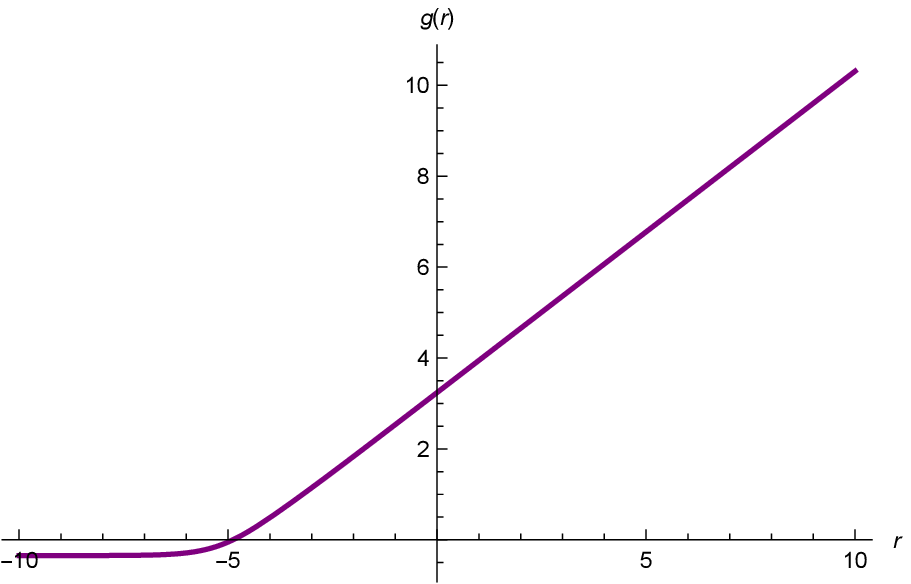}
                 \caption{Solution for $g$}
         \end{subfigure}\\

         ~ %add desired spacing between images, e. g. ~, \quad, \qquad, \hfill etc.
         \begin{subfigure}[b]{0.45\textwidth}
                 \includegraphics[width=\textwidth]{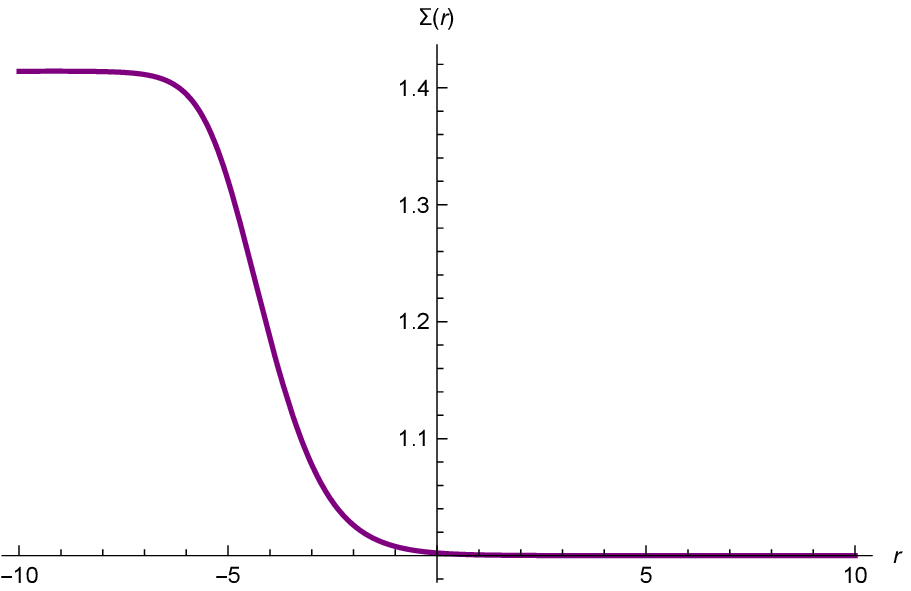}
                 \caption{Solution for $\Sigma$}
         \end{subfigure}\qquad 
         \begin{subfigure}[b]{0.45\textwidth}
                 \includegraphics[width=\textwidth]{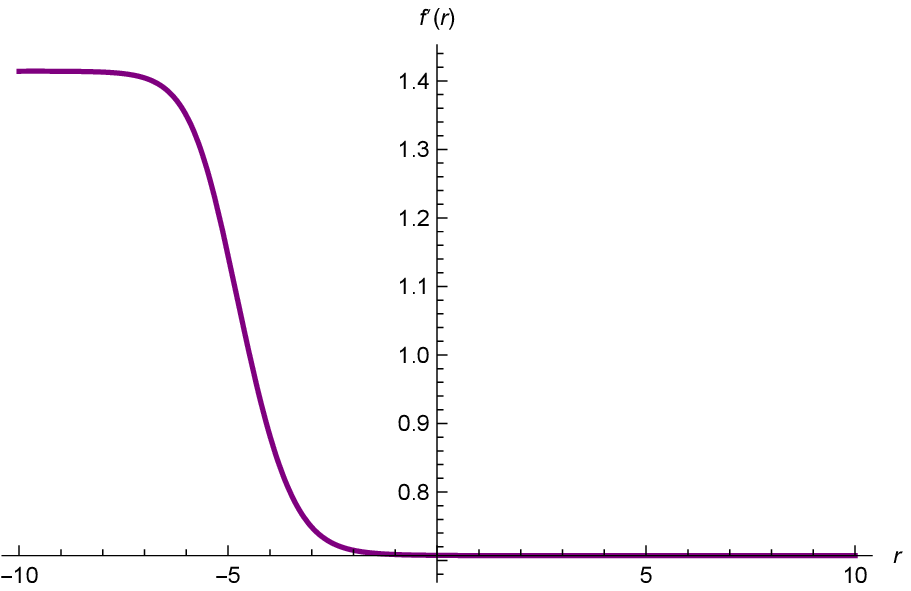}
                 \caption{Solution for $f'$}
         \end{subfigure}
         \caption{An RG flow from $AdS_5$ critical point with $U(1)\times SU(2)\times SU(2)$ symmetry to $AdS_3\times H^2$ critical point III for $g_1=1$ and $g_3=2g_1$.}\label{fig5}
\end{figure}

\indent Similarly, we can set $\phi_1=0$ and find a numerical solution interpolating between $AdS_5$ vacuum with $U(1)\times SU(2)\times SU(2)$ symmetry and $AdS_3\times H^2$ critical point III. The result is shown in figure \ref{fig5}. We have also verified that all of these solutions satisfy the corresponding field equations.

%%%%%%%%%%%%%%%%%%%%%%%%%%%%%%%%%%%%%%%%%%%%%%%%%%%%%%%%%%%%%%%%%%%%%%%%%%%%%%%%%%%%%%%%%%%%%%%%%%%%%%%%%%%%%%%%%%%%%%%%%%%%%%%%%%%%%%%%%
\section{Supersymmetric RG flows in $U(1)\times SO(3,1)$ gauge group}\label{U1_SO3_1_gauge_group}
In this section, we consider a non-compact gauge group $U(1)\times SO(3,1)$ with the embedding tensor
\begin{eqnarray}
\xi^{MN}&=&g_1(\delta^M_2\delta^N_1-\delta^M_1\delta^N_2),\\ 
f_{345} &=&f_{378} = -f_{468} = -f_{567} = -g_2\, .
\end{eqnarray} 
At the vacuum, the $U(1)\times SO(3,1)$ gauge group will be broken down to it maximal compact subgroup $U(1)\times SO(3)\subset U(1)\times SO(3,1)$. Under this unbroken symmetry, there is one scalar singlet from $SO(5,5)/SO(5)\times SO(5)$ corresponding to the non-compact generator
\begin{eqnarray}
Y = Y_{31} +Y_{42} - Y_{53}\, .
\end{eqnarray}
With the usual parametrization of the coset representative of the form
\begin{eqnarray}
L = e^{\phi Y},
\end{eqnarray}
the scalar potential is given by
\begin{eqnarray}
V&=&\frac{1}{16} \Sigma^{-2}e^{-6 \phi} g_2 \left[\left(1+3 e^{4 \phi}-16 e^{6 \phi}+3 e^{8 \phi}+e^{12 \phi}\right) g_2\phantom{\sqrt{2}}\right. \nonumber \\
& &\left.-4 \sqrt{2} e^{3 \phi} \left(1-3 e^{2 \phi}-3 e^{4 \phi}+e^{6 \phi}\right) g_1 \Sigma^3\right].\label{SO3_1_potential}
\end{eqnarray}
This potential admits only one $N=4$ supersymmetric $AdS_5$ vacuum due to the absence of flavor symmetry in the dual $N=2$ SCFT in agreement with the result of \cite{5D_N4_flow_Davide}. This critical point is located at
\begin{eqnarray}
\phi = 0 \qquad \textrm{and}\qquad  \Sigma =-\left(\frac{g_2}{\sqrt{2}g_1}\right)^{1/3}\, .\label{AdS5_SO3_1}
\end{eqnarray}   
As in the $U(1)\times SU(2)\times SU(2)$ gauge group, we can rescale $\Sigma$ such that $\Sigma=1$ at the $AdS_5$ vacuum. Equivalently, we can choose the value of $g_2$ to be $g_2=-\sqrt{2}g_1$. With this choice, the cosmological constant and $AdS_5$ radius are given by
\begin{equation}
V_0 = -3g_1^2\qquad \textrm{and}\qquad L^2 = -\frac{6}{V_0} = \frac{2}{g^2_1}\, .
\end{equation}
Scalar masses at this vacuum are given in table \ref{table3}. The spectrum is the same as that of the $N=4$ $AdS_5$ with $U(1)\times SU(2)_{\textrm{diag}}$ symmetry in the compact gauge group. Massless scalars in $\mathbf{3}_0$ representation are Goldstone bosons of the symmetry breaking $SO(3,1)\rightarrow SO(3)$.  
\begin{table}[h]
\centering
\begin{tabular}{|c|c|c|}
  \hline
  Scalar field representations & $m^2L^2\phantom{\frac{1}{2}}$ & $\Delta$  \\ \hline
  $\mathbf{1}_0$ & $-4$ &  $2$  \\
  $\mathbf{1}_0$ & $12$ &  $6$  \\
  $\mathbf{1}_{\pm 2}$ & $-3_{\times 4}$ &  $3$  \\
   $\mathbf{3}_{\pm 2}$ & $5_{\times 6}$ &  $5$  \\
    $\mathbf{3}_0$ & $-4_{\times 6}$ &  $2$  \\
      $\mathbf{3}_0$ & $0_{\times 3}$ &  $4$  \\
   $\mathbf{5}_0$ & $0_{\times 5}$ &  $4$  \\
  \hline
\end{tabular}
\caption{Scalar masses at the $N=4$ supersymmetric $AdS_5$ critical
point with $U(1)\times SO(3)$ symmetry and the
corresponding dimensions of the dual operators for the non-compact $U(1)\times SO(3,1)$ gauge group.}\label{table3}
\end{table}

\subsection{BPS equations and holographic RG flow solutions}
Since there is only one supersymmetric $AdS_5$ critical point, supersymmetric RG flows between $AdS_5$ critical points do not exist. We will look for solution describing a domain wall with one limit being the $AdS_5$ critical point identified above and another limit being a singular geometry dual to an $N=2$ non-conformal field theory.
\\
\indent With the same procedure as before, the superpotential in this case reads
\begin{eqnarray}
W = \frac{e^{-3 \phi} \left(1-3 e^{2 \phi}-3 e^{4 \phi}+e^{6 \phi}\right) g_2+2 \sqrt{2} g_1 \Sigma^3}{12 \Sigma}\, .
\end{eqnarray}
It can be easily verified that $W$ has only one critical point. The potential can be written in term of the superpotential as
\begin{eqnarray}
V = \frac{3}{2}\left[\Sigma^2\left(\frac{\partial W}{\partial \Sigma}\right)^2 + \left(\frac{\partial W}{\partial \phi}\right)^2\right] - 6W^2\, .
\end{eqnarray}
The BPS equations for this gauge group are given by
\begin{eqnarray}
\phi'= -\left(\frac{\partial W}{\partial \phi}\right)&=&\frac{e^{-3 \phi} \left(e^{2 \phi}-e^{4 \phi}+e^{6 \phi}-1\right) g_2}{4 \Sigma},\nonumber\\
\Sigma '= -\Sigma^2 \left(\frac{\partial W}{\partial \Sigma}\right) &=&\frac{1}{12} e^{-3 \phi} \left[4 \sqrt{2} e^{3 \phi} g_1 \Sigma^3-\left(1-3 e^{2 \phi}-3 e^{4 \phi}+e^{6 \phi}\right) g_2\right],\nonumber\\
A'=W &=& \frac{e^{-3 \phi} \left(1-3 e^{2 \phi}-3 e^{4 \phi}+e^{6 \phi}\right) g_2+2 \sqrt{2} g_1 \Sigma^3}{12 \Sigma}\, .
\end{eqnarray}
By combining these equations, the flow equations for the warp factor $A$ and the dilaton $\Sigma$ can be written as
\begin{eqnarray}
\Sigma'(\phi) &=&\frac{\Sigma \left[4 \sqrt{2} e^{3 \phi} g_1 \Sigma^3-g_2\left(1-3 e^{2 \phi}-3 e^{4 \phi}+e^{6 \phi}\right) \right]}{3g_2 \left(e^{2 \phi}-e^{4 \phi}+e^{6 \phi}-1\right)},\\
A'(\phi) &=& -\frac{\left(1-3 e^{2 \phi}-3 e^{4 \phi}+e^{6 \phi}\right) g_2+2 \sqrt{2} e^{3 \phi} g_1 \Sigma^3}{3 \left(e^{2 \phi}-e^{4 \phi}+e^{6 \phi}-1\right) g_2}\, .
\end{eqnarray}
The solution for $\Sigma$ can be readily obtained
\begin{eqnarray}\label{eq:Sigmaphi}
\Sigma= -\left[g_2\frac{e^{\phi} \left(e^{2 \phi}-1\right)}{g_2 C_1(1+e^{4 \phi})-\sqrt{2} g_1}\right]^{1/3}\, .
\end{eqnarray}
To make the flow approach the $AdS_5$ critical point, we choose the constant $C_1$ to be $C_1 = \frac{g_1}{\sqrt{2}g_2}$. This leads to 
\begin{eqnarray}\label{eq:Sigmaphi2}
\Sigma = -\left[\frac{\sqrt{2} g_2\,e^\phi}{g_1(1+ e^{2\phi})} \right]^{1/3}
\end{eqnarray}
which clearly gives $\Sigma=-\left(\frac{g_2}{\sqrt{2}g_1}\right)^{1/3}$ for $\phi=0$.
\\
\indent With the solution for $\Sigma(\phi)$, the solution for $A$ can be straightforwardly obtained. The result is 
\begin{eqnarray}
A = \frac{1}{6} \left[2 \phi +3 \ln\left(1-e^{2 \phi}\right)+\ln\left(1+e^{2 \phi}\right)-3 \ln\left(1+e^{4 \phi}\right)\right].
\end{eqnarray}
Finally, by redefining the radial coordinate $r$ to $\rho$ via $\frac{d\rho}{dr} = \Sigma^{-1}$, we find the solution for $\phi(\rho)$
\begin{eqnarray}
2g_2\rho&=& 2 \ln\left(1-e^{\phi}\right)-2 \ln\left(1+e^{\phi}\right) \nonumber \\
& &+\sqrt{2} \left[\ln\left(\sqrt{2} e^{\phi}+e^{2 \phi}+1\right)-\ln\left(\sqrt{2} e^{\phi}-e^{2 \phi}-1\right)\right]\label{phi_sol_non-compact}
\end{eqnarray}
where an additive integration constant has been discarded.
\\
\indent As $r\rightarrow \infty$, we find 
\begin{equation}
\Sigma\sim e^{-\frac{2r}{L}}\qquad \textrm{and}\qquad \phi\sim e^{\frac{2r}{L}}\, .
\end{equation}
The operator dual to $\phi$ is irrelevant as indicated by the value of $m^2L^2=12$ in table \ref{table3}. From the solution \eqref{phi_sol_non-compact}, $\phi\rightarrow \pm\infty$ at a finite value of $\rho$. Explicitly, we find that, as $\phi\rightarrow \pm\infty$,
\begin{equation}
\phi\sim \frac{1}{3}\ln\left[C+\frac{3g_2\rho}{4}\right]\qquad \textrm{and}\qquad \phi\sim -\ln\left[C-\frac{g_1\rho}{2}\right]
\end{equation}
for some constant $C$. In both cases, $\Sigma\rightarrow 0$ and $V\rightarrow \infty$. As a result, these singularities are unphysical by the criterion of \cite{Gubser_singularity}. 

\subsection{RG flows to $AdS_3\times \Sigma_2$ geometries}
\begin{figure}
         \centering
         \begin{subfigure}[b]{0.3\textwidth}
                 \includegraphics[width=\textwidth]{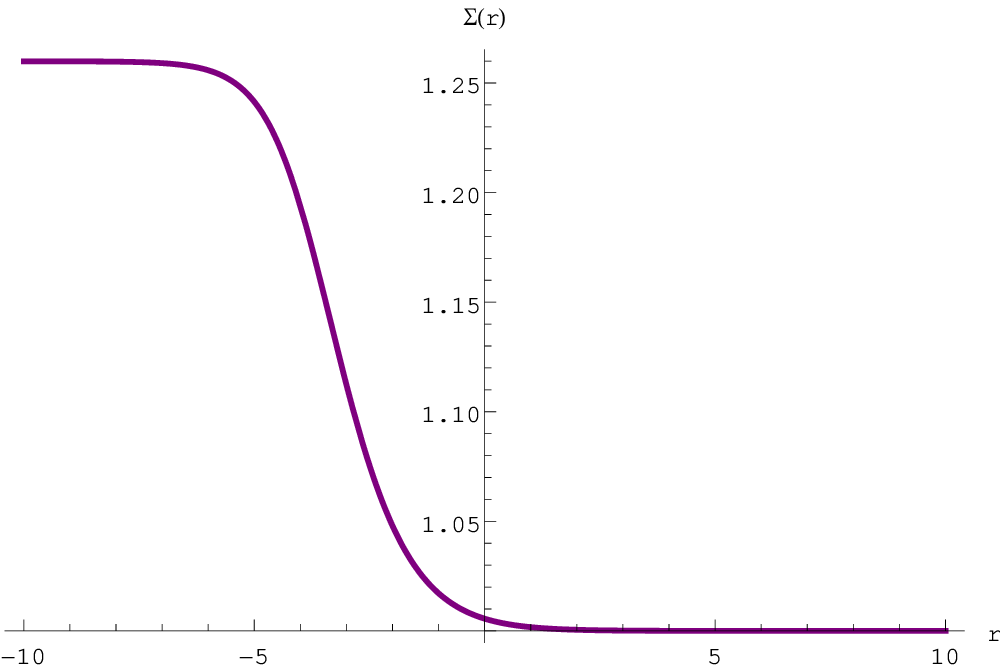}
                 \caption{Solution for $\Sigma$}
         \end{subfigure}\,\, 
\begin{subfigure}[b]{0.3\textwidth}
                 \includegraphics[width=\textwidth]{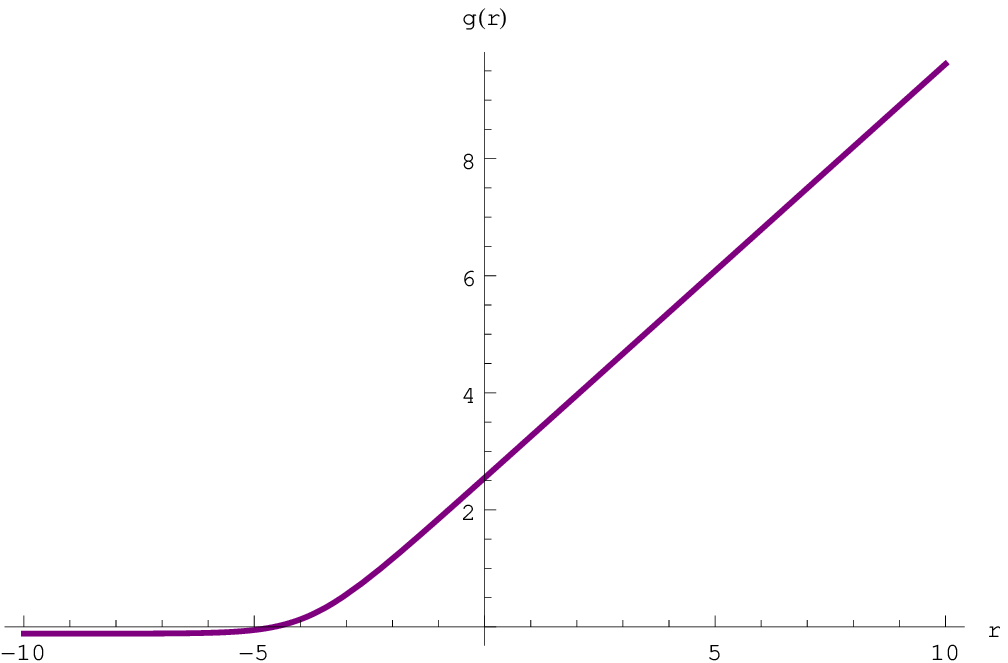}
                 \caption{Solution for $g$}
         \end{subfigure}\,\,
         ~ %add desired spacing between images, e. g. ~, \quad, \qquad, \hfill etc.
         \begin{subfigure}[b]{0.3\textwidth}
                 \includegraphics[width=\textwidth]{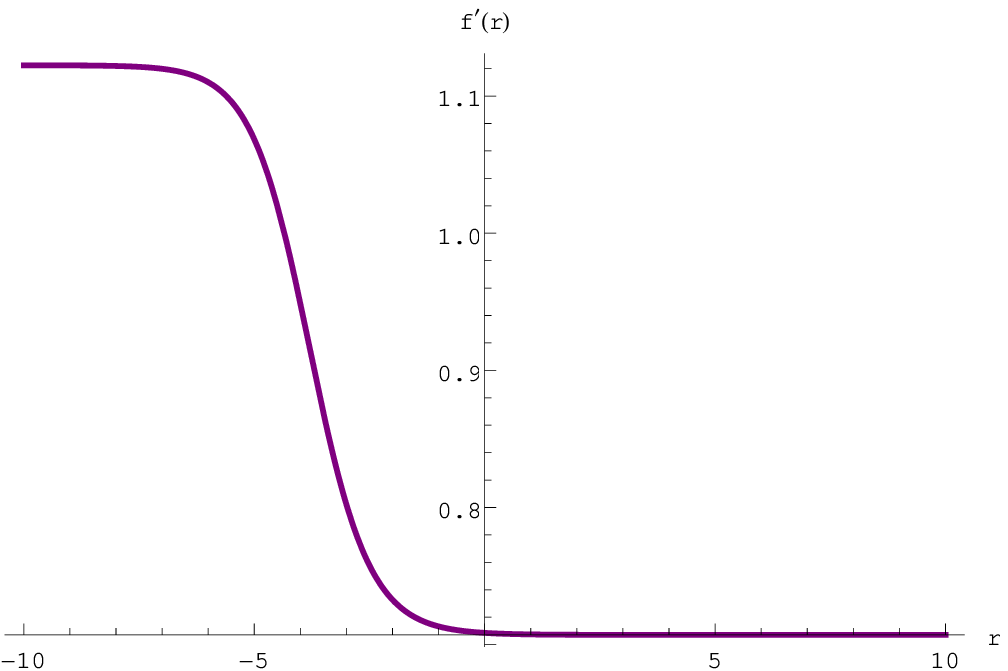}
                 \caption{Solution for $f'$}
         \end{subfigure}
         \caption{An RG flow from $N=4$ $AdS_5$ critical point with $U(1)\times SO(3)$ symmetry to $AdS_3\times H^2$ geometry in the IR from $U(1)\times SO(3,1)$ gauge group and $g_1=1$.}\label{fig6}
 \end{figure}

We now restrict ourselves to scalars which are invariant under $SO(2)\subset SO(3)\subset SO(3,1)$ whose generator is ${(T_5)_M}^N={f_{5M}}^N$. There are in total five singlets from $SO(5,5)/SO(5)\times SO(5)$. However, as in the case of $U(1)\times SU(2)\times SU(2)$ gauge group, we can truncate this set to just three singlets corresponding to the following non-compact generators
\begin{eqnarray}
\tilde{Y}_1 = Y_{31} + Y_{42}, \qquad
\tilde{Y}_2 = Y_{32}-Y_{41}, \qquad \tilde{Y}_3 = Y_{53}\, .
\end{eqnarray}
The coset representative is given by
\begin{eqnarray}
L = e^{\phi_1 \tilde{Y}_1}e^{\phi_2 \tilde{Y}_2}e^{\phi_3 \tilde{Y}_3},
\end{eqnarray}
and the potential reads 
\begin{eqnarray}
V&=&\frac{1}{16} \Sigma^{-2}\left[g^2_2 \left[\cosh(4 \phi_1-2 \phi_3)+\cosh(4 \phi_1+2\phi_3)\right.\phantom{\sqrt{2}} \right.\nonumber \\
& &\left.+\cosh(2 \phi_3) \left(6+4 \cosh(4 \phi_2) \sinh^2(2 \phi_1)\right)\right] -16g_2^2
\nonumber\\ 
& & \left. +16 \Sigma^3\sqrt{2} g_1 g_2\left[\cosh\phi_3+\cosh(2 \phi_2) \sinh(2 \phi_1) \sinh(\phi_3)\right] \right]
\end{eqnarray}
which admits only a single supersymmetric critical point at which all vector multiplet scalars vanish. 
\\
\indent The metric ansatz is still given by \eqref{metric_ansatz_AdS3}. We will consider the twists obtained from turning on $U(1)\times U(1)\subset U(1)\times SO(3,1)$ gauge fields along $\Sigma_2$. These gauge fields will be denoted by $A^0$ and $A^5$. As in the previous section, the twists from $A^0$ and $A^5$ cannot be turned on simultaneously. Furthermore, the $A^0$ twist does not lead to $AdS_3\times \Sigma_2$ solutions. We will therefore consider only the twist from $A^5$. It turns out that the two-form fields can also be consistently set to zero provided that we set the gauge fields $A^1=A^2=0$.
\\
\indent With the same ansatz as in \eqref{vector_ansatz}, together with the projectors \eqref{gamma_r_projector} and \eqref{gamma_theta_phi_projector}, we find the following BPS equations after using the twist condition $g_2a_5=1$  
\begin{eqnarray}
f'&=& -\frac{1}{24 \Sigma}e^{-2 \phi_1-2 \phi_2-\phi_3-2 g} \left[e^{2 g} \left(1-e^{4 \phi_1}+e^{4 \phi_2}+4 e^{2 (\phi_1+\phi_2)}-e^{4 (\phi_1+\phi_2)}-e^{2 \phi_3}
\right.\right.\nonumber\\
&&\left.\left.+4 e^{2 (\phi_1+\phi_2+\phi_3)}+e^{4 \phi_1+2 \phi_3}-e^{4 \phi_2+2 \phi_3}
+e^{4 \phi_1+4 \phi_2+2 \phi_3}\right) g_2
\right.\nonumber\\
&&\left.-4\kappa a_5 e^{2 (\phi_1+\phi_2)} \left(1+e^{2 \phi_3}\right) \Sigma^2-4 \sqrt{2} e^{2 \phi_1+2 \phi_2+\phi_3+2 g} g_1 \Sigma^3\right],\\
%\end{eqnarray}
%\begin{eqnarray}
g'&=& -\frac{1}{24 \Sigma}e^{-2 \phi_1-2 \phi_2-\phi_3-2 g} \left[e^{2 g} \left(1-e^{4 \phi_1}+e^{4 \phi_2}+4 e^{2 (\phi_1+\phi_2)}-e^{4 (\phi_1+\phi_2)}-e^{2 \phi_3}\right.\right.\nonumber\\
&&\left.\left.+4 e^{2 (\phi_1+\phi_2+\phi_3)}+e^{4 \phi_1+2 \phi_3}-e^{4 \phi_2+2 \phi_3}
+e^{4 \phi_1+4 \phi_2+2 \phi_3}\right) g_2
\right.\nonumber\\
&&\left.+8\kappa a_5 e^{2 (\phi_1+\phi_2)} \left(1+e^{2 \phi_3}\right) \Sigma^2 - 4 \sqrt{2} e^{2 \phi_1+2 \phi_2+\phi_3+2 g} g_1 \Sigma^3\right], \\
%\end{eqnarray}
%\begin{eqnarray}
\Sigma '&=& -\frac{1}{24} e^{-2 \phi_1-2 \phi_2-\phi_3-2 g} \left[e^{2 g} \left(1-e^{4 \phi_1}+e^{4 \phi_2}+4 e^{2 (\phi_1+\phi_2)}-e^{4 (\phi_1+\phi_2)}-e^{2 \phi_3}
\right.\right.\nonumber\\
&&\left.\left.+4 e^{2 (\phi_1+\phi_2+\phi_3)}+e^{4 \phi_1+2 \phi_3}-e^{4 \phi_2+2 \phi_3}+e^{4 \phi_1+4 \phi_2+2 \phi_3}\right) g_2
\right.\nonumber\\
&&\left.-4\kappa a_5 e^{2 (\phi_1+\phi_2)} \left(1+e^{2 \phi_3}\right) \Sigma^2 + 8 \sqrt{2} e^{2 \phi_1+2 \phi_2+\phi_3+2 g} g_1 \Sigma^3\right],\\
%\end{eqnarray}
%\begin{eqnarray}
\phi_1'&=& \frac{e^{-2 \phi_1+2 \phi_2-\phi_3} \left(1+e^{4 \phi_1}\right) \left(e^{2 \phi_3}-1\right) g_2}{2 \left(1+e^{4 \phi_2}\right) \Sigma}, \\
\phi_2'&=& \frac{e^{-2 \phi_1-2 \phi_2-\phi_3} \left(e^{4 \phi_1}-1\right) \left(e^{4 \phi_2}-1\right) \left(e^{2 \phi_3}-1\right) g_2}{8 \Sigma},\\
\phi_3'&=& \frac{1}{8 \Sigma}e^{-2 \phi_1-2 \phi_2-\phi_3-2 g} \left[e^{2 g} \left(e^{4 \phi_1}-e^{4 \phi_2}-4 e^{2 (\phi_1+\phi_2)}+e^{4 (\phi_1+\phi_2)}-e^{2 \phi_3}-1
\right.\right.\nonumber\\
& &\left.+4 e^{2 (\phi_1+\phi_2+\phi_3)}+e^{4 \phi_1+2 \phi_3}-e^{4 \phi_2+2 \phi_3}+e^{4 \phi_1+4 \phi_2+2 \phi_3}\right) g_2\nonumber\\
& &\left.+4\kappa a_5 e^{2 (\phi_1+\phi_2)} \left(e^{2 \phi_3}-1\right) \Sigma^2\right].
\end{eqnarray}
Unlike the compact gauge group considered in the previous section, the above equations admit only one $AdS_3\times H^2$ solution given by  
\begin{eqnarray}
\phi_1 &=& \phi_2 = \phi_3 = 0, \qquad  \Sigma = -\left(\frac{\sqrt{2}\,g_2}{g_1}\right)^{1/3}, \nonumber\\
g &=& -\frac{1}{2}\ln\left[\frac{1}{a_5}\left(\frac{g^2_1 g_2}{2}\right)^{1/3}\right],\qquad L_{AdS_3} = \left(\frac{\sqrt{2}}{g_1 g_2^2}\right)^{1/3} \, .
\end{eqnarray}
\indent To find an RG flow solution interpolating between this $AdS_3\times H^2$ and the $AdS_5$ critical point \eqref{AdS5_SO3_1}, we can consistently set all $\phi_i$, $i=1,2,3$, to zero and $\kappa=-1$. The remaining BPS equations read
\begin{eqnarray}
f'&=& -\frac{2 g_2+2 a_5 e^{-2 g} \Sigma^2-\sqrt{2} g_1 \Sigma^3}{6 \Sigma},
\nonumber\\
g'&=& -\frac{2 g_2-4 a_5 e^{-2 g} \Sigma^2-\sqrt{2} g_1 \Sigma^3}{6 \Sigma},
\nonumber\\
\Sigma '&=& -\frac{1}{3} \left(g_2+a_5 e^{-2 g} \Sigma^2+\sqrt{2} g_1 \Sigma^3\right).
\end{eqnarray}
A numerical solution to these equations is shown in figure \ref{fig6}. Similar to an analogous solution in the compact gauge group, this solution is the same as the universal RG flow considered in \cite{flow_acrossD_bobev} since it does not involve scalars from vector multiplets.
%%%%%%%%%%%%%%%%%%%%%%%%%%%%%%%%%%%%%%%%%%%%%%%%%%%%%%%%%%%%%%%%%%%%%%%%%%%%%%%%%%%%%%%%%%%%%%%%%%%%%%%%%%%%%%%%%%%%%%%%%%%%%%%%%%%%%%%%%
\section{Conclusions and discussions}\label{conclusion}
We have studied gauged $N=4$ supergravity in five dimensions coupled to five vector multiplets with compact and non-compact gauge groups $U(1)\times SU(2)\times SU(2)$ and $U(1)\times SO(3,1)$. For $U(1)\times SU(2)\times SU(2)$ gauge group, we have recovered two supersymmetric $N=4$ $AdS_5$ vacua with $U(1)\times SU(2)\times SU(2)$ and $U(1)\times SU(2)_{\textrm{diag}}$ symmetries together with the RG flow interpolating between them found in \cite{5D_N4_flow_Davide}. However, we have also given the full mass spectra for scalar fields at both critical points which have not been studied in \cite{5D_N4_flow_Davide}. These should be useful in the holographic context since it provides information about dimensions of operators dual to the supergravity scalars. For $U(1)\times SO(3,1)$ gauge group, there is only one $N=4$ supersymmetric $AdS_5$ critical point with vanishing vector multiplet scalars. We have given an RG flow solution from an $N=2$ SCFT dual to this vacuum to a non-conformal field theory dual to a singular geometry. However, this singularity is unphysical within the framework of $N=4$ gauged supergravity. It would be interesting to embed this solution in ten or eleven dimensions and further investigate whether this singularity is resolved or has any physical interpretation in the context of string/M-theory. 
\\
\indent We have also considered $AdS_3\times \Sigma_2$ solutions by turning on gauge fields along $\Sigma_2$. We have found that in order to preserve eight supercharges, the twists from the $U(1)$ factor in the gauge group and the Cartan $U(1)\subset SU(2)$, denoted by the parameters $a_0$ and $a_5$, cannot be performed simultaneously. It should also be noted that for less supersymmetric solutions, both $a_0$ and $a_5$ can be non-vanishing such as $\frac{1}{4}$-BPS solutions found in \cite{5D_N4_Romans} for pure $N=4$ gauged supergravity with $U(1)\times SU(2)$ gauge group. It would also be interesting to look for more general solutions of this type. 
\\
\indent For $U(1)\times SU(2)\times SU(2)$ gauge group, we have identified a number of $AdS_3\times H^2$ solutions preserving eight supercharges. We have given numerical RG flow solutions from the two $AdS_5$ vacua to these $AdS_3\times H^2$ geometries. For $U(1)\times SO(3,1)$ gauge group, there is one $AdS_3\times H^2$ solution when all scalars from vector multiplets vanish. The solution preserves eight supercharges similar to the solutions in the compact gauge group. A numerical RG flow between this  solution and the $N=4$ $AdS_5$ vacuum has also been given. All of these solutions describe twisted compactifications of $N=2$ SCFTs on $H^2$ and should be of interest in holographic studies of $N=2$ SCFTs in four dimensions and in the context of supersymmetric black strings. It is noteworthy that the space of $AdS_5$ and $AdS_3$ solutions in the compact gauge group is much richer than that of the non-compact gauge group. This is in line with similar studies of half-maximal gauged supergravities in other dimensions.
\\
\indent There are a number of future works extending our results presented here. It is interesting to consider flow solutions with non-vanishing two-form fields similar to the recently found solutions in seven and six dimensions in \cite{7D_sol_Dibitetto,7D_SUSY_solutions,6D_SUSY_solutions_GD}. These solutions will also  give a description of conformal defects in the dual $N=2$ SCFTs. Furthermore, finding Janus solution within this $N=4$ gauged supergravity is also of particular interest. This can be done by an analysis similar to that initiated in \cite{curved_DW_Dallagata1} and \cite{curved_DW_Dallagata2}. Up to now, this type of solutions has only appeared in $N=8$ and $N=2$ gauged supergravities, see for example \cite{5D_Janus_CK,5D_Janus_Suh}.     
\vspace{0.5cm}\\
%%%%%%%%%%%%%%%%%%%%%%%%%%%%%%%%%%%%%%%%%%%%%%%%%%%%%%%%%%%%%%%%%%%%%%%%%%%%%%%%%%%%%%%%%%%%%%%%%%%%%%%%%%%%%%%%%%%%%%%%%%%%%%%%%%%%%%%%%
{\large{\textbf{Acknowledgement}}} \\
P. K. is supported by The Thailand Research Fund (TRF) under grant RSA5980037.
%%%%%%%%%%%%%%%%%%%%%%%%%%%%%%%%%%%%%%%%%%%%%%%%%%%%%%%%%%%%%%%%%%%%%%%%%%%%%%%%%%%%%%%%%%%%%%%%%%%%%%%%%%%%%%%%%%%%%%%%%%%%%%%%%%%%%%%%%%%%%%%%%%%%%%%%%%%%%%%%

\end{document}